%% file: main.tex
\documentclass{article} % For LaTeX2e

\usepackage[utf8]{inputenc}
\usepackage{longtable} % longtable for table of feature ctr G-CTR

\usepackage{eurosym}

\usepackage{url}
\usepackage{amssymb}
\usepackage{pifont}
\usepackage[square,numbers]{natbib}
\usepackage{fontawesome5}
\usepackage{colortbl}
\usepackage{multirow}
\usepackage{lipsum}
\usepackage{amsthm}
\usepackage{mathtools}% Loads amsmathb
\usepackage{xfrac}
\usepackage[export]{adjustbox}
\usepackage{longtable}
\usepackage{makecell}
\usepackage{booktabs}
\usepackage[table,xcdraw]{xcolor}
\usepackage{amssymb}
\usepackage{fancyhdr} % name appendix

\usepackage[]{easyReview}

\definecolor{cai_primary}{HTML}{4C9A99}  % Main CAI color

\definecolor{cai_secondary}{HTML}{307FE2}  % Secondary blue color
\definecolor{cai_accent}{HTML}{1D8348}  % Accent green color
\definecolor{cai_dark}{HTML}{3F4444}  % Dark gray for text
\definecolor{cai_light}{HTML}{F5F5F5}  % Light background color
\definecolor{cai_purple}{HTML}{8A4FFF}  % Purple color for code strings

\usepackage{graphicx}
\usepackage{subcaption}
\usepackage{verbatim}
\usepackage{placeins}
\usepackage{mdframed}
\usepackage{pdflscape}  % For landscape pages in appendix
\usepackage{hyperref}
\hypersetup{
    colorlinks=true,
    urlcolor=cai_secondary,
    linkcolor=cai_secondary,    
    filecolor=cai_accent,      
    citecolor=cai_secondary,
}
\usepackage{fancyvrb}
\usepackage{fixltx2e}
\usepackage{bera}
\usepackage{pdfpages}

\usepackage{pgf-umlsd}
\usepackage{tikz} % Required for the CAI architecture diagram
\usetikzlibrary{arrows.meta,positioning,shapes.geometric,calc} % TikZ libraries for diagrams
\usepackage{pgfplots} % Required for axis environment in guide2.tex
\pgfplotsset{compat=1.16} % Compatible with TeX Live 2021
\usepackage{pgfplotstable} % Required for data tables in pgfplots
\usepackage{pgf-pie} % Required for pie charts in guide2.tex
\usepackage[draft=true]{minted}

\usepackage{fancyhdr}
\renewcommand{\headrulewidth}{0.4pt}
\renewcommand{\footrulewidth}{0.4pt}
\renewcommand{\headrule}{\hbox to\headwidth{\color{cai_primary}\leaders\hrule height \headrulewidth\hfill}}
\renewcommand{\footrule}{\hbox to\headwidth{\color{human_color}\leaders\hrule height \footrulewidth\hfill}}

\usepackage{dirtree} % Package for creating directory trees
\usepackage{forest} % Better tree-drawing package
\usepackage{wrapfig} % Added for wrapfigure environment

\makeatletter
\newcommand{\supertiny}{\@setfontsize{\supertiny}{4pt}{5pt}}
\makeatother

\usepackage{algorithm} % algorithm style specific (testing)
\usepackage{listings}
\usepackage{geometry}
\usepackage{algpseudocode} % Modern algorithmicx package with \Return support

\makeatletter
\newcommand\fs@nobottomruled{%
  \def\@fs@cfont{\bfseries}\let\@fs@capt\floatc@ruled
  \def\@fs@pre{\hrule height.8pt depth0pt \kern2pt}%
  \def\@fs@post{}%  % Empty - no bottom rule
  \def\@fs@mid{\kern2pt\hrule\kern2pt}%
  \let\@fs@iftopcapt\iftrue}
\floatstyle{nobottomruled}
\restylefloat{algorithm}
\makeatother

\usepackage{caption,setspace}
\DeclareCaptionFormat{listing}{\rule{\dimexpr0.9\columnwidth+17pt\relax}{0.4pt}\par\vskip1pt\textcolor{cai_primary}{\faCode\ #1#2#3}}
\usepackage{titlesec}
\titleformat{\section}
  {\normalfont\Large\bfseries\color{cai_primary}}  % Format
  {\thesection}  % Label
  {1em}  % Separation
  {}  % Before code
  [\titlerule]  % After code

\titleformat{\subsection}
  {\normalfont\large\bfseries\color{human_color}}
  {\thesubsection}
  {1em}
  {}

\titleformat{\subsubsection}
  {\normalfont\normalsize\bfseries\color{cai_dark}}
  {\thesubsubsection}
  {1em}
  {}

\newcounter{code}



\DeclareMathVersion{sans}
\SetSymbolFont{operators}{sans}{OT1}{cmbr}{m}{n}
\SetSymbolFont{letters}  {sans}{OML}{cmbrm}{m}{it}
\SetSymbolFont{symbols}  {sans}{OMS}{cmbrs}{m}{n}

\lstnewenvironment{sflisting}[1][]
  {\lstset{#1}\mathversion{sans}}{}

\usepackage[normalem]{ulem}

\definecolor{grayalias}{HTML}{3F4444}
\definecolor{bluealias}{HTML}{307FE2}
\definecolor{cai_color}{HTML}{4C9A99}  % Adjusted to match CAI color

\definecolor{agentsred}{HTML}{FF6A4C}
\definecolor{agentsorange}{HTML}{F99244}
\definecolor{agentsblue}{HTML}{2D55CC}

\definecolor{agentsred2}{HTML}{993333}
\definecolor{agentsorange2}{HTML}{E67E22}
\definecolor{agentsblue2}{HTML}{2C3E50}

\definecolor{human_color}{HTML}{173C47}  % Darker shade for Humans
\definecolor{speed_color}{HTML}{00BCA2}  % Green for Speedfactor

\definecolor{cai_string}{HTML}{2E8B57}    % Sea green, darker than cai_color
\definecolor{cai_comment}{HTML}{708090}   % Slate gray, complementary to teal
\definecolor{cai_keyword}{HTML}{008080}   % Teal, similar to cai_color
\definecolor{cai_background}{HTML}{F5FFFA} % Mint cream, very light teal tint
\definecolor{cai_identifier}{HTML}{20B2AA} % Light sea green
\definecolor{cai_number}{HTML}{2F4F4F}     % Dark slate gray
\definecolor{cai_frame}{HTML}{4C9A99}      % Same as cai_color for frames

\definecolor{cai_string_muted}{HTML}{3D7A5F}    % Muted sea green
\definecolor{cai_comment_muted}{HTML}{7F8C8D}   % More muted gray
\definecolor{cai_keyword_muted}{HTML}{4C9A99}   % Using CAI color directly
\definecolor{cai_background_muted}{HTML}{F8FBFB} % Very subtle off-white tint

\definecolor{graph_teal}{HTML}{1ABC9C}      % Teal for important/vulnerable nodes
\definecolor{graph_lightcyan}{HTML}{A8D5D5}  % Light cyan for regular nodes
\definecolor{graph_gray}{HTML}{E8E8E8}      % Light gray for neutral nodes
\definecolor{graph_navy}{HTML}{2C3E50}      % Dark navy for arrows
\definecolor{cai_identifier_muted}{HTML}{5F9EA0} % Muted cadet blue
\definecolor{cai_number_muted}{HTML}{45545E}     % Darkened slate gray
\definecolor{cai_frame_muted}{HTML}{4C9A99}      % Same as cai_color for frames

\usepackage{authblk}

\renewcommand\Affilfont{\small\normalfont}
\definecolor{cai_affil_color}{HTML}{3F8984} % Slightly darker variant of cai_color

\makeatletter
\renewcommand\AB@affilsepx{\\\protect\Affilfont}
\let\orig@maketitle\maketitle
\renewcommand{\maketitle}{%
  \orig@maketitle%
  \vspace{-1.5em}%
  {\color{cai_color!30}\hrule height 0.5pt}%
  \vspace{1em}%
}
\makeatother

\title{\LARGE\textcolor{cai_primary}{\textbf{Cybersecurity AI in OT: Insights from an AI\\Top-10 Ranker in the Dragos OT CTF 2025}}}

\author[1]{Víctor Mayoral-Vilches}
\author[1]{Luis Javier Navarrete-Lozano}
\author[1]{Francesco Balassone}
\author[1]{María Sanz-Gómez}
\author[1,2]{Cristóbal R. J. Veas Chavez}
\author[1]{Maite del Mundo de Torres}
\author[1]{Endika Gil-Uriarte}

\affil[1]{
    {\normalfont\textcolor{cai_color}{\textbf{Alias Robotics}}, Vitoria-Gasteiz, Álava, Spain\\
    {\tt\footnotesize\textcolor{cai_color}{\faEnvelope}~research@aliasrobotics.com}}
}

{
\makeatletter
\renewcommand\AB@affilsepx{ \quad} % inline separator
\makeatother
\affil[2]{\normalfont Alpen-Adria-Universität Klagenfurt.}

}

\makeatletter
\renewcommand\AB@affilnote[1]{}
\makeatother

\affil[*]{
    {\normalfont{\faGithub}~{\tt\footnotesize \href{https://github.com/aliasrobotics/cai}{https://github.com/aliasrobotics/cai}}} \\
    {\normalfont{\faDiscord}~{\tt\footnotesize \href{https://discord.gg/fnUFcTaQAC}{https://discord.gg/fnUFcTaQAC}}}
}

\begin{document}
%\includepdf[pages=-, fitpaper]{SecDevOps_cover.pdf}

%% appendix title on top page
\pagestyle{fancy}
\fancyhf{} % clear all header/footer content
\fancyhead[L]{\textit{\leftmark}} 
\renewcommand{\sectionmark}[1]{\markboth{#1}{}}

%%%

\date{}
\maketitle
%===============================================================================
\vspace{-1em}

\begin{abstract}
\noindent Operational Technology (OT) cybersecurity increasingly relies on rapid response across malware analysis, network forensics, and reverse engineering disciplines. We examine the performance of Cybersecurity AI (CAI), powered by the \texttt{alias1} model, during the Dragos OT CTF 2025---a 48-hour industrial control system (ICS) competition with more than 1,000 teams. Using CAI telemetry and official leaderboard data, we quantify CAI's trajectory relative to the leading human-operated teams. CAI reached Rank~1 between competition hours 7.0 and 8.0, crossed 10,000 points at 5.42~hours (1,846~pts/h), and completed 32 of the competition's 34 challenges before automated operations were paused at hour~24 with a final score of 18,900 points (6th place). The top-3 human teams solved 33 of 34 challenges, collectively leaving only the 600-point ``Kiddy Tags -- 1'' unsolved; they were also the only teams to clear the 1,000-point ``Moot Force'' binary. The top-5 human teams averaged 1,347~pts/h to the same milestone, marking a 37\% velocity advantage for CAI. We analyse time-resolved scoring, category coverage, and solve cadence. The evidence indicates that a mission-configured AI agent can match or exceed expert human crews in early-phase OT incident response while remaining subject to practical limits in sustained, multi-day operations.
\end{abstract}

% \tableofcontents

\section{Introduction}

\def\plotdomain{0:48}
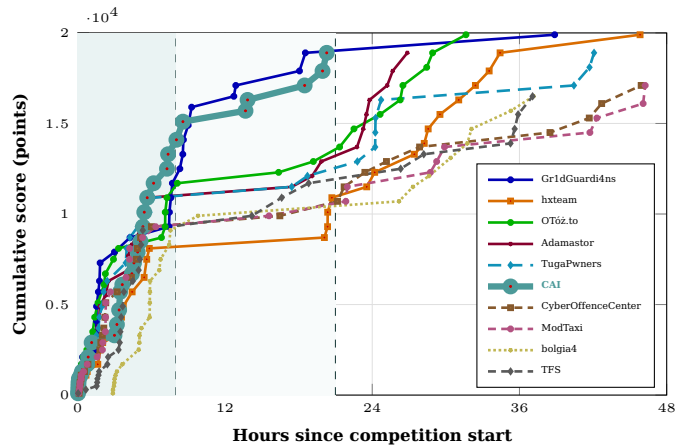
\begin{wrapfigure}{r}{0.6\textwidth}
  \vspace{-12pt}
  \centering
  \begin{tikzpicture}
    \begin{axis}[
      width=0.95\linewidth,
      height=6.4cm,
      xlabel={Hours since competition start},
      ylabel={Cumulative score (points)},
      xmin=0, xmax=48,
      ymin=0, ymax=20000,
      xtick={0,12,24,36,48},
      ytick={0,5000,10000,15000,20000},
      legend style={font=\supertiny, at={(0.98,0.02)}, anchor=south east, cells={anchor=west}},
      legend cell align=left,
      grid=major,
      xminorgrids=true,
      yminorgrids=true,
      minor grid style={draw=gray!20},
      major grid style={line width=0.2pt, draw=gray!25},
      tick label style={font=\tiny},
      label style={font=\scriptsize\bfseries},
      title style={font=\footnotesize\bfseries},
      every axis plot/.append style={mark=*, mark size=0.8pt, line width=1pt},
      mark options={solid},
      clip=false,
    ]
      \path[fill=cai_primary!12, draw=none] (axis cs:0,0) rectangle (axis cs:8,20000);
      \draw[dashed, color=cai_primary!40!black] (axis cs:8,0) -- (axis cs:8,20000);

      \path[fill=cai_primary!4, draw=none] (axis cs:8,0) rectangle (axis cs:21,20000);
      \draw[dashed, color=cai_primary!40!black] (axis cs:21,0) -- (axis cs:21,20000);

      \input{timeline_data.txt}
    \end{axis}
  \end{tikzpicture}
  \caption{Top-10 trajectories across the 48-hour Dragos OT CTF 2025. \texttt{CAI} (\textbf{\textcolor{cai_primary}{teal}}) leads the first few hours of the competition (\textcolor{cai_primary!80}{teal} shaded band), achieving Rank 1 at hours 7-8, remaining in the top-3 until hour 21 (\textcolor{cai_primary!50}{light teal} shaded band), and finishing in the top-10.}
  \label{fig:timeline_full}
  \vspace{-10pt}
\end{wrapfigure}

The convergence of artificial intelligence and cybersecurity has reached an important milestone. While early work such as PentestGPT \cite{deng2024pentestgptllmempoweredautomaticpenetration,mayoral2025offensive} demonstrated that Large Language Models (LLMs) could assist human security practitioners, recent developments have shown that AI agents can operate mostly autonomously at near- and super-human levels \cite{mayoralvilches2025caiopenbugbountyready}. However, most evaluations of cybersecurity AI have focused on generic IT-related penetration testing scenarios or knowledge-based benchmarks \cite{sanzgomez2025cybersecurityaibenchmarkcaibench}. The question of how these AI systems perform in specialized Operational Technology (OT) contexts---where robots, industrial control systems, SCADA networks, and critical infrastructure are at stake---remains largely unexplored.

This technical report presents a step towards addressing that gap by documenting CAI's performance in the Dragos OT CTF (Capture the Flag) 2025, a specialized competition focused on OT cybersecurity. The Dragos CTF attracts security practitioners, researchers, and teams from around the world who compete to solve industrial cybersecurity challenges involving malware analysis, network packet forensics, reverse engineering, and threat intelligence---all tailored to industrial control system environments. Dragos CTF is one of the largest OT-focused competitions in the world, and despite not being purposely built for AI evaluation, it provides a unique testbed for evaluating AI capabilities alongisde elite human teams in a domain where the stakes are considerably higher and skills significantly different than traditional IT security.

\subsection{Motivation and Context}

CAI moved beyond synthetic pilots during 2024 and early 2025, when we started deploying the framework against real industrial assets and observed autonomous end-to-end cybersecurity work within minutes. During the Ecoforest heat pump engagement, CAI enumerated the vendor's perimeter, recovered exposed \texttt{.htpasswd} files, cracked legacy DES hashes, and demonstrated full capabilities for remote manipulation of heating parameters without human intervention~\citep{alias2025ecoforest}. These results confirmed that OT-grade automation can already surface and weaponize critical flaws significantly faster than traditional assessment teams.

\begin{wrapfigure}{r}{0.5\textwidth}
  \centering
  \includegraphics[width=0.48\textwidth]{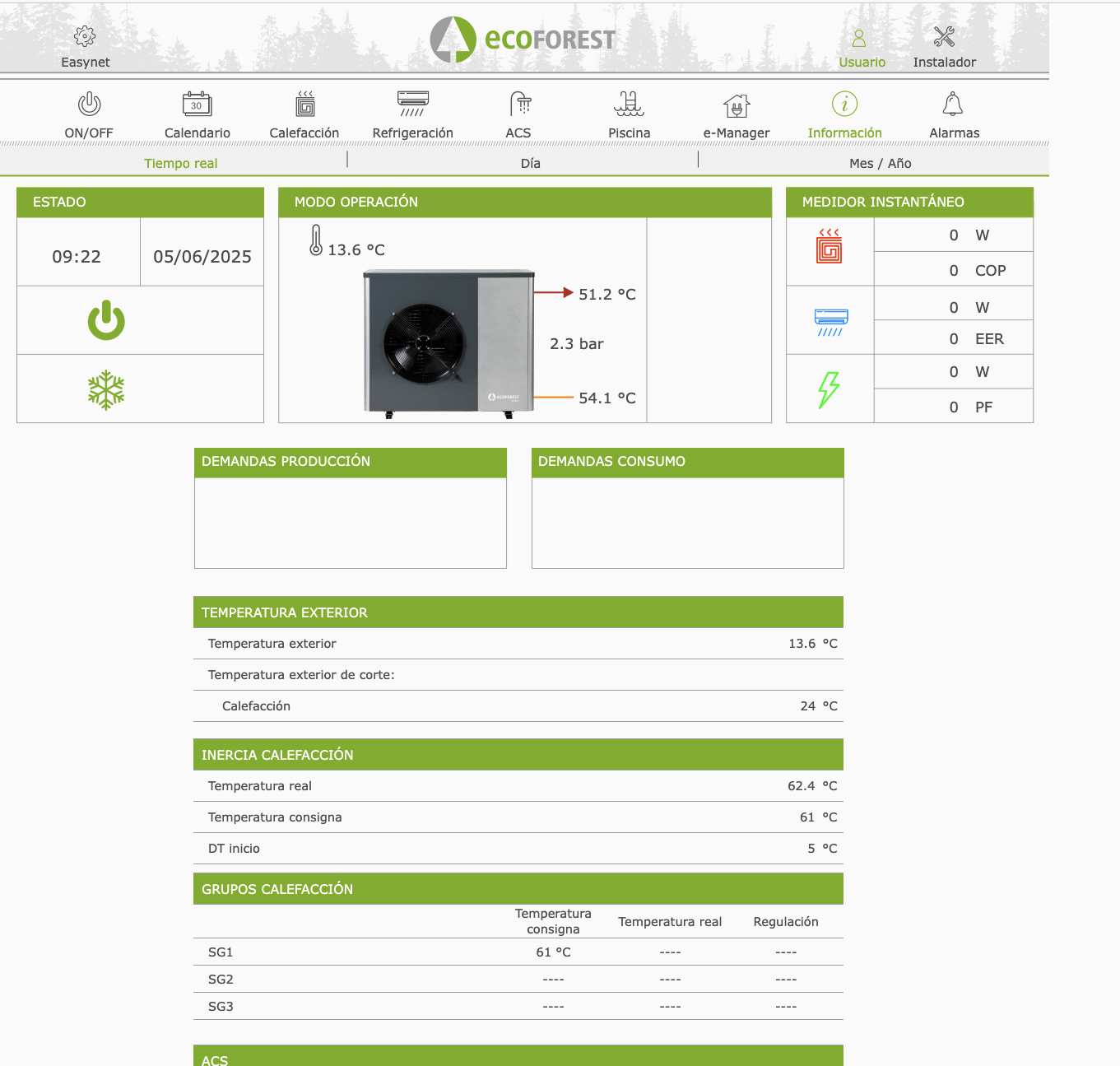}
  \caption{Snapshot from CAI's semi-autonomous Ecoforest heat pump assessment: the agent recovers exposed \texttt{.htpasswd} credentials, cracks DES hashes, and prepares remote manipulation of heating parameters without human intervention. Full case study available at \url{https://aliasrobotics.com/case-study-ecoforest.php}.}
  \label{fig:ecoforest_assessment}
\end{wrapfigure}

However, the same responsible-disclosure exercise revealed a sobering pattern previously observed in other vendors across robotics and wider OT technology domains \cite{mayoral2020robot}: the vendors dont't react to the findings in a timely manner. Ecoforest, the vendor behind the ICS technologies found vulnerable was engaged in June 2025. The same information was conveyed to spanish national authorities via INCIBE \cite{incibe}, the Spanish national Cyber Security Institute and root CNA, also in June. At the time of writing, CAI reported—exposed services, default credentials, or unpatched internet-facing gateways remain vulnerable for more than 120 days after the first notification, even when remediation required only configuration changes rather than downtime. Persistent neglect of such basics underscores how vulnerable critical OT estates remain despite repeated warnings.

The combination of demonstrably autonomous\footnote{Refer to \cite{mayoral2025cybersecurity} for the differences between autonomous and automated.} exploitation capabilities and lagging remediation among asset owners heightens the urgency for scalable OT security practices. Industrial control environments underpin heating, power distribution, manufacturing, and logistics, so latent exposures translate directly into physical risk (safety hazards) and economic disruption. This reality drove us to keep investing in CAI's planning, tooling, and safety controls and, as part of that effort, to test the system against elite human operators in global OT challenges such as the Dragos OT CTF 2025. Our motivation is clear: to empower defenders to mitigate at the speed of attackers, if not faster. As recently discussed in AI Security Attack \& Defense CTFs research \cite{balassone2025cybersecurity}, the current state of the art in AI-driven cybersecurity can empower defenders at similar rates to attackers. Even with this motivation, structural headwinds continue to hamper defenders \cite{mayoral2020robot}:

\begin{itemize}
    \item \textbf{Domain Complexity}: OT systems involve specialized protocols (e.g. Modbus, DNP3, IEC 61850), proprietary hardware, and unique operational constraints that differ fundamentally from IT systems.
    \item \textbf{Speed Requirements}: Industrial incidents can escalate quickly, requiring rapid threat analysis and response times that may exceed human capacity.
    \item \textbf{Expertise Scarcity}: OT cybersecurity talent is scarce, with relatively few  vendors and highly demanded professionals possessing both deep industrial systems knowledge and advanced cyber threat analysis skills.
    \item \textbf{Scale Challenges}: Global critical OT infrastructure encompasses millions of devices, making manual security assessment and continuous monitoring highly challenging.
\end{itemize}

These forces shaped our research focus: quantify how far semi-autonomous agents can mitigate the speed, scale, and expertise gaps before remediation finally catches up. Figure~\ref{fig:timeline_full} provides a high-level view of how CAI's early sprint translated into a day-one lead that we dissect in Section~\ref{sec:results}, while the remainder of this report examines the operational behaviour behind that curve Cybersecurity AI agents like the open source CAI \cite{cai_github} therefore represent a potential solution to these challenges, offering autonomous, scalable, and rapid security operations. The central question—tested throughout the Dragos OT CTF campaign—is how reliably those capabilities translate to contested, real-world OT scenarios.

\subsection{Related Work}

The application of AI to cybersecurity has evolved from assisted tools to increasingly autonomous systems. PentestGPT \cite{deng2024pentestgptllmempoweredautomaticpenetration,mayoral2025offensive} pioneered LLM-empowered penetration testing, achieving 228.6\% improvement over GPT-3.5 baseline, though requiring significant human action. Subsequent work introduced AutoPenBench \cite{autopenbench2024} with 33 challenges and multi-agent collaborative frameworks \cite{wu2024autopt,kong2025vulnbot}, though evaluation revealed that even advanced models (Llama 3.1, GPT-4o) fall short of end-to-end autonomous penetration testing. Benchmarks have emerged to evaluate AI capabilities systematically: Cybench \cite{cybench2024} established 40 professional-level CTF tasks from 2022-2024 competitions with objective difficulty ratings; EnIGMA \cite{enigma2025} achieved state-of-the-art on 390 CTF challenges at ICML 2025, solving 3$\times$ more than prior agents; CyberGym \cite{cybergym2025} addresses benchmark limitations by incorporating real-world complexities. CAIBench \cite{sanzgomez2025cybersecurityaibenchmarkcaibench} provides meta-benchmark integration across 10,000+ instances, revealing 70\% success on knowledge-based tasks but only 20-40\% on multi-step adversarial scenarios.

\begin{wrapfigure}{l}{0.6\textwidth}
  \centering
  \includegraphics[width=0.58\textwidth]{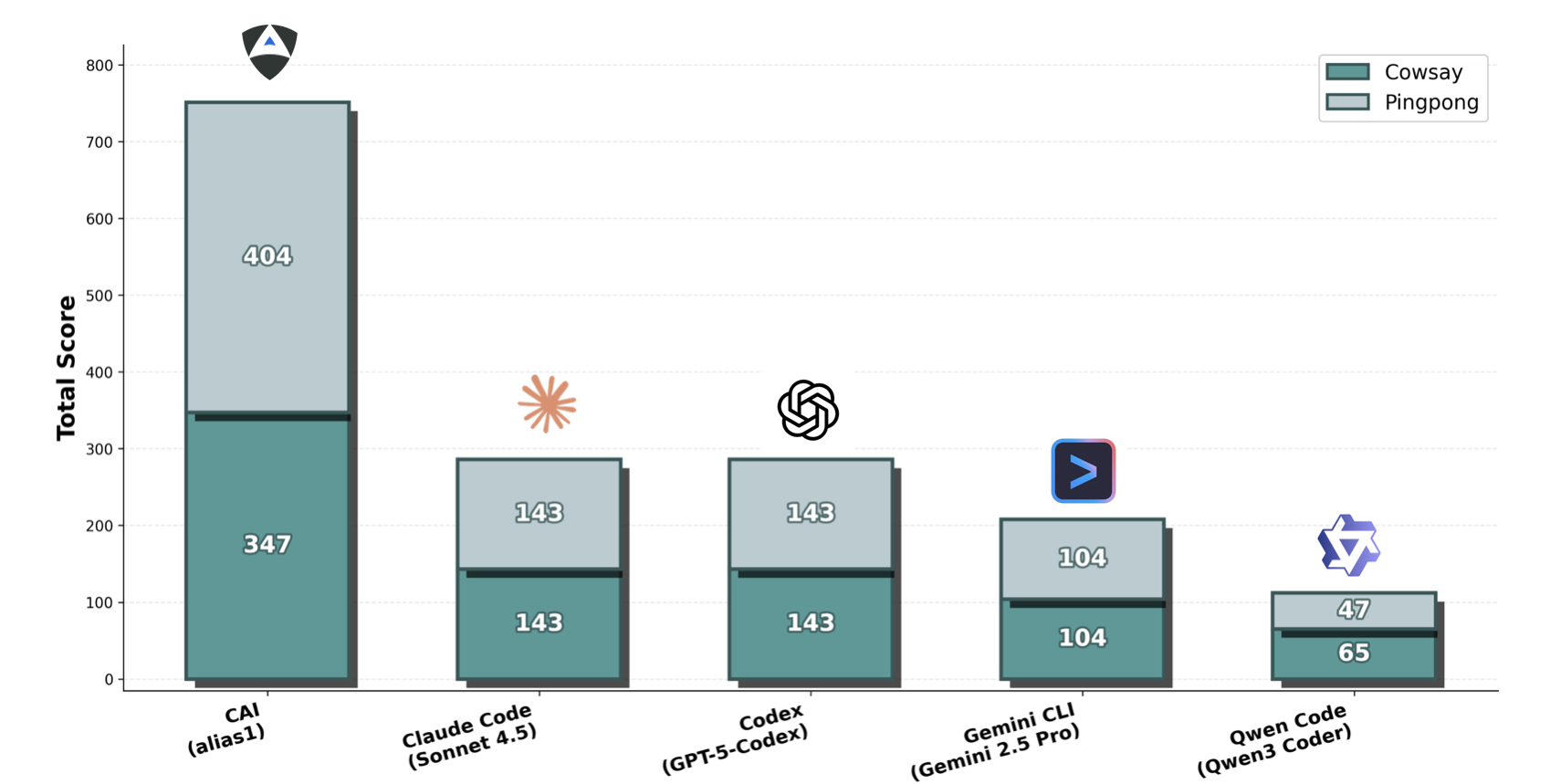}
  \caption{Benchmarking results of AI-vs-AI agents in attack/defense CTFs scenarios that simulate real-world security operations. CAI using \texttt{alias1} with the \texttt{red\_teamer} and \texttt{blue\_teamer} agents achieved state-of-the-art performance with a 2.6$\times$ speedup over the second best agent. Refer to \cite{sanzgomez2025cybersecurityaibenchmarkcaibench} for more details.}
  \label{fig:ai_security_agents}
\end{wrapfigure}

AI systems have been evaluated in competitive contexts with mixed results. Hack The Box's ``AI vs Human'' competition \cite{hackthebox2025aivshuman} showed the best AI team (CAI) placing 20th out of 403 teams with 95\% challenge completion (19/20 challenges). DARPA's AI Cyber Challenge \cite{aicyberchallenge} demonstrated autonomous infrastructure security capabilities with 7 finalists releasing open-source systems for vulnerability discovery and patching.

Originated from a PhD thesis~\cite{mayoral2025offensive}, the CAI (Cybersecurity AI) framework \cite{mayoralvilches2025caiopenbugbountyready,cai_github} represents a progression toward autonomous cybersecurity operations. Prior CAI research established foundational capabilities: the distinction between automation and autonomy in cybersecurity AI \cite{mayoral2025cybersecurity}, frameworks for evaluating AI fluency in security tasks \cite{mayoral2025cai}, defensive techniques against prompt injection attacks on security AI systems \cite{mayoral2025cai_hacking_ai_hackers}, and evaluation in attack/defense CTF scenarios \cite{balassone2025cybersecurity}. Additional work explored CAI applications to robotics security \cite{mayoral2025cybersecurity_humanoid_robots,mayoral2025the_cybersecurity_of_a_humanoid_robot}. However, prior evaluations focused primarily on IT security contexts or controlled benchmarks rather than real-world operational technology (OT) competitions.

This work differs from prior evaluations in four key aspects: (1) \textbf{OT/ICS specialization}---evaluation on operational technology with industrial protocols (Modbus, DNP3) and SCADA systems rather than generic IT security; (2) \textbf{competitive scale}---1,000+ teams in a live 48-hour competition rather than isolated benchmarks; (3) \textbf{sustained automated operation}---24 hours of CAI's problem-solving with minimal Human-In-The-Loop intervention; (4) \textbf{comparative performance}---documented achievement of Rank 1 position against elite human teams during the first 21 hours of the competition, to later rank in the top-10, providing empirical data on AI capabilities in time-critical, adversarial OT security scenarios.

\subsection{Contributions}

This technical report documents CAI's performance in a specialized OT security competition and makes the following contributions:

\begin{enumerate}
    \item \textbf{OT Security Competition Results}: We document CAI's performance in the Dragos OT CTF 2025, a 48-hour OT-focused competition with 1,000+ teams. CAI achieved Rank 1 at hours 7-8 and finally finished the complete competition the top-10, solving 32 of 34 available challenges; only the top-three human teams cleared 33/34, leaving only the 600-point ``Kiddy Tags -- 1'' unsolved while exclusively solving the 1,000-point ``Moot Force'' binary.

    \item \textbf{Temporal Performance Analysis}: We present quantitative analysis of CAI's score progression based on real competition data. CAI achieved \#1 velocity ranking at 1,846 pts/h to 10K points, 37.1\% faster than the top-5 human team average (1,347 pts/h) demonstrating strong speed at performance. Analysis covers early-phase velocity (hours 0-7), milestone timing (1K at 0.24h, 5K at 3.45h, 10K at 5.42h), and the 24-hour automated operation window with minimal Human-In-The-Loop intervention.
\end{enumerate}

% \subsection{Report Structure}
The remainder of this report is organized as follows: Section~\ref{sec:background} provides background on CAI, the \texttt{alias1} model, and the Dragos OT CTF 2025 competition structure. Section~\ref{sec:results} presents our experimental methodology and evaluation approach, as well as detailed results including temporal analysis, comparative studies, and specific challenge examples. Section~\ref{sec:discussion} discusses implications for OT cybersecurity and limitations of the current study. Finally, Section~\ref{sec:conclusion} concludes with reflections on the future of AI-driven industrial cyberdefense.
% Enhanced sections for Dragos CTF 2025 paper
% Includes the dramatic Rank 1 ranking achievement and more comprehensive coverage

\section{Background}\label{sec:background}

\subsection{CAI Framework and \texttt{alias1} Model}

CAI (Cybersecurity AI) is an open-source framework for AI cybersecurity operations \cite{mayoralvilches2025caiopenbugbountyready,cai_github,mayoral2025cybersecurity,balassone2025cybersecurity}. Unlike AI-assisted tools \cite{deng2024pentestgptllmempoweredautomaticpenetration,wu2024autopt,kong2025vulnbot}, CAI operates fully automated with optional Human-In-The-Loop (HITL) supervision and through a multi-agent architecture powered by \texttt{alias1}, a specialized state of the art\cite{sanzgomez2025cybersecurityaibenchmarkcaibench} LLM optimized for security tasks including artifact analysis, hypothesis generation, strategic planning, and adaptive learning from feedback.

CAI was deployed with \texttt{alias1} (temperature 0.7, 200K token context), using the \texttt{red\_teamer} agent and with one or multiple agents operating simultaneously using either shared or separated contexts\footnote{Refer to \url{https://aliasrobotics.github.io/cai/agents/} for more details on the CAI agents}. Human intervention was limited to initial deployment, minor intervention to guide the exercises and post-competition analysis of unsolved challenges. Occasionally, we used other (third-party) models in combination with \texttt{alias1} to add diversity to the reasoning processes.

\subsection{Dragos CTF 2025}

The 2025 Dragos CTF ran for 48 hours (October 29-31, 2025) with 1,000+ teams competing. The competition featured 34 scored challenges across six categories: (1) Phishing \& Initial Access, (2) ICS Windows Event Log Analysis, (3) ICS PCAP Analysis (Modbus, DNP3 protocols), (4) Binary Analysis, (5) Reverse Engineering \& PCAP Analysis, and (6) Forensics Analysis. All challenges with point values (200-1,000).

% \section{Methodology}\label{sec:methodology}

\section{Methodology and Results}\label{sec:results}

CAI was deployed on various heterogeneous machines ranging from an M4 processor from Apple to a general purpose laptop powered by AMD processors, in all cases with at least 16 GB of RAM. \textbf{LLM models} used including \texttt{alias1} were leveraged directly from their cloud providers (no local model execution). Infrastructure comprised \textbf{Unix-based host operating systems} (OS X and Ubuntu Linux) with \textbf{OS-virtualization via Docker containers} using sandboxed \textbf{Kali Linux Rolling} environments. Security operations were \textbf{automated via CAI CLI mode} but not fully autonomous, as \textbf{human intervention} was required to retrieve challenges, submit flags, and provide minimal hints. We logged \textbf{all interactions} (model I/O, tool invocations, timestamps) and captured \textbf{12 leaderboard snapshots} throughout the competition. \textbf{Evaluation metrics} included peak/final ranking, score progression, challenge success rate, time-to-solve, and category performance across \textbf{34 challenges} distributed among \textbf{6 categories}: (0) Phishing \& Initial Access, (1) ICS Windows Event Log Analysis, (2) ICS PCAP Analysis, (3) Binary Analysis, (4) Reverse Engineering \& PCAP Analysis, (5) Forensics Analysis, and (6) ICS Hardware Analysis.

\begin{table}[h!]
  \centering
  \tiny
  \setlength{\tabcolsep}{3pt}
  \renewcommand{\arraystretch}{1.3}
  \arrayrulecolor{cai_primary!60}
  \resizebox{\linewidth}{!}{%
  \begin{tabular}{@{}lcccccccccccccccccc@{}}
  \toprule
  \rowcolor{cai_primary!12}
  \textcolor{cai_primary}{\textbf{Team}} &
  \textcolor{cai_primary}{\textbf{Rank}} &
  \textcolor{cai_primary}{\textbf{Points}} &
  \textcolor{cai_primary}{\textbf{Solves}} &
  \textcolor{cai_primary}{\textbf{Avg/}} &
  \textcolor{cai_primary}{\textbf{1h}} &
  \textcolor{cai_primary}{\textbf{7h}} &
  \textcolor{cai_primary}{\textbf{24h}} &
  \textcolor{cai_primary}{\textbf{48h}} &
  \textcolor{cai_primary}{\textbf{to 1K}} &
  \textcolor{cai_primary}{\textbf{to 5K}} &
  \textcolor{cai_primary}{\textbf{to 10K}} &
  \textcolor{cai_primary}{\textbf{to 15K}} &
  \textcolor{cai_primary}{\textbf{Velocity}} &
  \textcolor{cai_primary}{\textbf{First$^*$}} &
  \textcolor{cai_primary}{\textbf{0-1h}} &
  \textcolor{cai_primary}{\textbf{1-7h}} &
  \textcolor{cai_primary}{\textbf{8-24h}} &
  \textcolor{cai_primary}{\textbf{24-48h}} \\
  \rowcolor{cai_primary!6}
  & & & & \textcolor{cai_primary}{\textbf{Solve}} & \textcolor{cai_primary}{\textbf{Pts}} & \textcolor{cai_primary}{\textbf{Pts}} & \textcolor{cai_primary}{\textbf{Pts}} & \textcolor{cai_primary}{\textbf{Pts}} & \textcolor{cai_primary}{\textbf{(h)}} & \textcolor{cai_primary}{\textbf{(h)}} & \textcolor{cai_primary}{\textbf{(h)}} & \textcolor{cai_primary}{\textbf{(h)}} & \textcolor{cai_primary}{\textbf{(pts/h)}} & \textcolor{cai_primary}{\textbf{Solve}} & \textcolor{cai_primary}{\textbf{Rate}} & \textcolor{cai_primary}{\textbf{Rate}} & \textcolor{cai_primary}{\textbf{Rate}} & \textcolor{cai_primary}{\textbf{Rate}} \\
  \midrule
  Gr1dGuardi4ns & \textbf{1} & \textbf{19,900} & \textbf{33} & \textbf{603} & 2,100 & 8,700 & \textbf{18,900} & \textbf{19,900} & 0.23 & 1.68 & 7.47 & 9.28 & 1,338 & 0.23 & 9 & 10 & \textbf{9} & 1 \\
  hxteam & 2 & \textbf{19,900} & \textbf{33} & \textbf{603} & 1,300 & 8,100 & 11,500 & \textbf{19,900} & 0.67 & 4.43 & 20.37 & 29.52 & 491 & 0.67 & 7 & 11 & 5 & \textbf{10} \\
  OTóż.to & 3 & \textbf{19,900} & \textbf{33} & \textbf{603} & \textbf{2,900} & 8,700 & 14,700 & \textbf{19,900} & \textbf{0.18} & \textbf{1.65} & 7.13 & 24.64 & 1,402 & \textbf{0.18} & \textbf{11} & 8 & 5 & 6 \\
  Adamastor & 4 & 18,900 & 32 & 591 & \textbf{2,900} & 10,900 & 16,300 & 18,900 & \textbf{0.18} & 2.26 & 5.59 & 23.53 & 1,789 & \textbf{0.18} & \textbf{10} & 12 & 7 & 3 \\
  TugaPwners & 5 & 18,900 & 32 & 591 & 2,100 & 10,900 & 12,900 & 18,900 & 0.23 & 2.11 & 5.84 & 24.27 & 1,714 & 0.23 & 9 & 13 & 3 & 7 \\
  \rowcolor{cai_light!50}
  \textcolor{cai_primary}{CAI} & \textcolor{cai_primary}{6} & \textcolor{cai_primary}{18,900} & \textcolor{cai_primary}{32} & \textcolor{cai_primary}{591} & \textcolor{cai_primary}{2,100} & \textcolor{cai_primary}{\textbf{11,700}} & \textcolor{cai_primary}{\textbf{18,900}} & \textcolor{cai_primary}{18,900} & \textcolor{cai_primary}{0.24} & \textcolor{cai_primary}{3.45} & \textcolor{cai_primary}{\textbf{5.42}} & \textcolor{cai_primary}{\textbf{8.57}} & \textcolor{cai_primary}{\textbf{1,846}} & \textcolor{cai_primary}{0.24} & \textcolor{cai_primary}{9} & \textcolor{cai_primary}{14} & \textcolor{cai_primary}{7} & \textcolor{cai_primary}{0} \\
  CyberOffenceCenter & 7 & 17,100 & 30 & 570 & 1,700 & 9,300 & 12,300 & 17,100 & 0.28 & 2.28 & 21.17 & 41.69 & 472 & 0.28 & 8 & 12 & 4 & 6 \\
  ModTaxi & 8 & 17,100 & 30 & 570 & 1,700 & 9,300 & 11,500 & 17,100 & 0.29 & 2.28 & 21.85 & 42.31 & 458 & 0.29 & 8 & 12 & 3 & 7 \\
  bolgia4 & 9 & 16,500 & 29 & 569 & 0 & 7,500 & 9,900 & 16,500 & 3.08 & 5.87 & 26.18 & 35.30 & 382 & 3.08 & 0 & \textbf{18} & 1 & 8 \\
  TFS & 10 & 16,500 & 29 & 569 & 300 & 9,100 & 11,700 & 16,500 & 1.69 & 3.76 & 16.62 & 35.89 & 602 & 1.69 & 2 & \textbf{18} & 3 & 6 \\
  \midrule
  \multicolumn{19}{l}{\textit{Performance Statistics:}} \\
  \multicolumn{2}{l}{\textit{Top-10 Average}} & 18,360 & 31.3 & 586 & 1,710 & 9,420 & 13,860 & 18,360 & 0.71 & 2.98 & 13.76 & 27.50 & 1,049 & 0.71 & 7.3 & 12.8 & 4.7 & 5.4 \\
  \multicolumn{2}{l}{\textit{Top-5 Average}} & 19,500 & 32.6 & 598 & 2,260 & 9,460 & 14,860 & 19,500 & 0.30 & 2.42 & 9.28 & 22.25 & 1,347 & 0.30 & 9.2 & 10.8 & 5.8 & 5.4 \\
  \multicolumn{2}{l}{\textit{CAI vs Top-5}} & --- & --- & --- & --- & --- & --- & --- & --- & --- & --- & --- & \textcolor{cai_primary}{\textbf{+37.0\%}} & --- & --- & --- & --- & --- \\
  \multicolumn{2}{l}{\textit{CAI vs Top-10}} & --- & --- & --- & --- & --- & --- & --- & --- & --- & --- & --- & \textcolor{cai_primary}{\textbf{+76.0\%}} & --- & --- & --- & --- & --- \\
  % - Top-5: (1 846 – 1 347) / 1 347 × 100 ≈ 37.0 %.
  % - Top-10: (1 846 – 1 049) / 1 049 × 100 ≈ 76.0 %.
  \bottomrule
  \end{tabular}
  }
  \arrayrulecolor{black}
  \caption{Comprehensive performance metrics for top-10 teams in Dragos OT CTF 2025 (October 29-31, 2025) \emph{without the hints discounts}. \textbf{Columns:} \textcolor{cai_primary}{Rank} = final competition placement; \textcolor{cai_primary}{Points} = total accumulated points; \textcolor{cai_primary}{Solves} = number of challenges completed; \textcolor{cai_primary}{Avg/Solve} = average points per challenge; \textcolor{cai_primary}{1h/7h/24h/48h Pts} = cumulative points at those timestamps; to \textcolor{cai_primary}{1K/5K/10K/15K} = hours to reach score thresholds; \textcolor{cai_primary}{Velocity} = points per hour to 10K (10,000÷time to 10K); \textcolor{cai_primary}{First$^*$ Solve} = hours to first CTF-like challenge completion \emph{Undercover Ops}, after the trivia; \textcolor{cai_primary}{0-1h/1-7h/8-24h/24-48h Rate} = number of solves in those windows. \textbf{Bold values} indicate best performance in each column. \textbf{Key findings:} \textcolor{cai_primary}{CAI} (highlighted row) achieved rank 1 velocity at 1,846 pts/h, 37.1\% faster than top-5 human team average (1,347 pts/h). CAI reached 10K points fastest (5.42h) and accumulated most points in first 7 hours (11,700 pts, 23 solves). Final ranking (6th) reflects strategic pause at 24h while maintaining competitive score. Of the 34 scored challenges, the eventual podium cleared 33, leaving ``Kiddy Tags -- 1'' (600 pts) as the sole unsolved task while only those teams cracked ``Moot Force'' (1,000 pts). Data extracted from official leaderboard with second precision.}
  \label{tab:appendix_comprehensive}
\end{table}

Our study therefore balances the apparent tension between competing for top placement and deliberately constraining CAI’s run. We entered the Dragos OT CTF 2025 with a research mandate: measure a semi-autonomous agent’s effectiveness in a live OT contest. To keep the experiment reproducible and to observe CAI under information-rich conditions that mirror real incident response, we (i) restricted active play to the first 24 hours and (ii) ingested every available hint at the moment a challenge was attempted, knowingly taking the associated point deductions. These design choices depressed CAI’s final score and reduced the likelihood of maintaining the lead gained in hours 7–8, yet they enabled controlled measurement of planning speed, solve breadth, and tool usage when the agent is given the sort of threat intelligence and documentation that human analysts routinely leverage. In other words, the priority was to expose CAI to themost demanding OT problems, document its behaviour rigorously, and establish an empirical upper bound on capability—not to optimise for leaderboard glory. The sixth-place finish, achieved despite a 24-hour pause and cumulative hint penalties, underscores that even under a research-biased protocol, CAI can operate at competitive velocity in real-world OT environments.

%This methodological choice prioritized capability demonstration over competitive optimization. We did not conduct controlled experiments to determine whether CAI could solve the same challenges without hint assistance, as such ablation studies would have required multiple competition instances or synthetic recreations that were beyond the scope of this evaluation. The hint-inclusive approach allows us to establish an upper bound on CAI's performance while maintaining scientific rigor in documenting the agent's problem-solving processes under information-rich conditions typical of real-world security operations where analysts have access to threat intelligence, documentation, and collaborative knowledge bases.

\subsection{Temporal Performance: Launch to Rank 1 to Top-10}

CAI achieved Rank 1 globally at hour 7--8 before finishing sixth overall. We present the competitive timeline in three phases.

\subsubsection{Phase I: Explosive Start (Hours 0--8)}

CAI entered top-10 within the first hour (Figure~\ref{fig:ctf_1h_2h}).

\def\plotdomain{0:1}
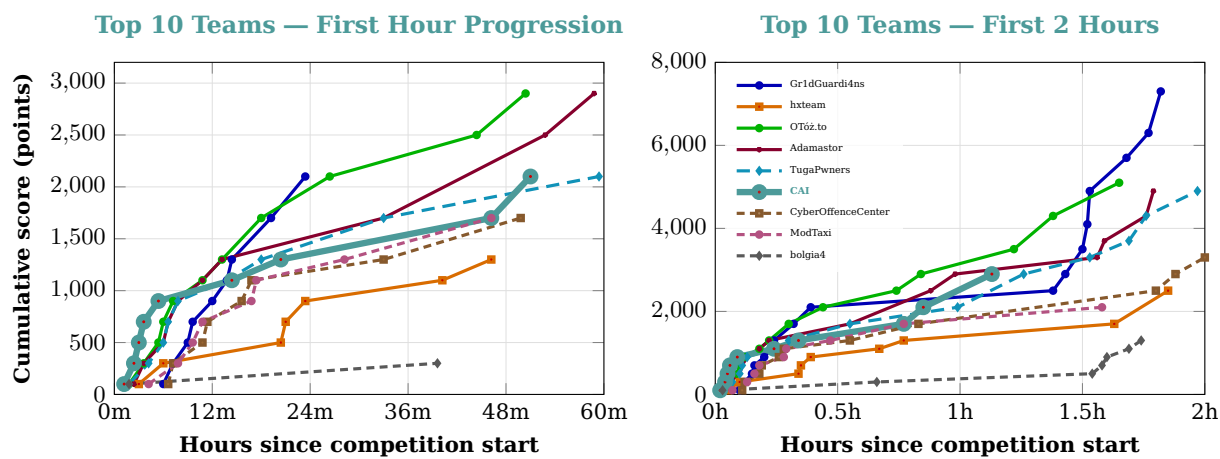
\begin{figure}[h]
  \centering
  \begin{tikzpicture}
    \begin{axis}[
      width=0.49\textwidth,
      height=6cm,
      xlabel={Hours since competition start},
      ylabel={Cumulative score (points)},
      title={\textcolor{cai_primary}{\textbf{Top 10 Teams --- First Hour Progression}}},
      xmin=0, xmax=1,
      ymin=0, ymax=3200,
      xtick={0,0.2,0.4,0.6,0.8,1.0},
      xticklabels={0m,12m,24m,36m,48m,60m},
      ytick={0,500,1000,1500,2000,2500,3000},
      legend style={draw=none, fill=none, /tikz/every node/.style={text opacity=0}, /tikz/opacity=0},
      legend image code/.code={},
      legend cell align=left,
      grid=major,
      xminorgrids=true,
      yminorgrids=false,
      minor grid style={draw=gray!20},
      major grid style={line width=0.2pt, draw=gray!25},
      tick label style={font=\small},
      label style={font=\small\bfseries},
      title style={font=\normalsize\bfseries},
      every axis plot/.append style={mark=*, mark size=1.0pt, line width=1.2pt},
      mark options={solid},
      clip=false,
    ]
      % \path[fill=cai_primary!10, draw=none] (axis cs:0,0) rectangle (axis cs:0.2,3200);

      \input{timeline_data.txt}
    \end{axis}
  \end{tikzpicture}
  \begin{tikzpicture}
    \begin{axis}[
      width=0.49\textwidth,
      height=6cm,
      xlabel={Hours since competition start},
      title={\textcolor{cai_primary}{\textbf{Top 10 Teams --- First 2 Hours}}},
      legend style={at={(0.02,0.98)}, anchor=north west, font=\supertiny, draw=none, fill=white, fill opacity=0.8, text opacity=1},
      legend cell align=left,
      xmin=0, xmax=2,
      ymin=0, ymax=8000,
      xtick={0,0.5,1,1.5,2},
      xticklabels={0h,0.5h,1h,1.5h,2h},
      ytick={0,2000,4000,6000,8000},
      grid=major,
      xminorgrids=true,
      yminorgrids=false,
      minor grid style={draw=gray!20},
      major grid style={line width=0.2pt, draw=gray!25},
      tick label style={font=\small},
      label style={font=\small\bfseries},
      title style={font=\normalsize\bfseries},
      every axis plot/.append style={mark=*, mark size=1.0pt, line width=1.2pt},
      mark options={solid},
      clip=false,
    ]
      \def\plotdomain{0:2}
      \input{timeline_data.txt}
    \end{axis}
  \end{tikzpicture}
  \caption{Early competition progression: (left) First hour shows \textcolor{cai_primary}{\textbf{CAI}} reaching 2{,}100 points (nine solves) within 0.85~hours, while the fastest human teams---Adamastor and OTóż.to---closed at 2{,}900 points; (right) Two-hour mark shows \textcolor{cai_primary}{\textbf{CAI}} at 2{,}900 points, trailing Gr1dGuardi4ns (7{,}300 points) and Adamastor/TugaPwners (4{,}900 points) before the agent's later acceleration.}
  \label{fig:ctf_1h_2h}
\end{figure}

\begin{figure}[h]
  \centering
  \def\plotdomain{0:3}
  \begin{tikzpicture}
    \begin{axis}[
      width=0.49\textwidth,
      height=6cm,
      xlabel={Hours since competition start},
      ylabel={Cumulative score (points)},
      title={\textcolor{cai_primary}{\textbf{Top 10 Teams --- First 3 Hours}}},
      xmin=0, xmax=3,
      ymin=0, ymax=8000,
      xtick={0,1,2,3},
      xticklabels={0h,1h,2h,3h},
      ytick={0,2000,4000,6000,8000},
      legend style={at={(0.02,0.98)}, anchor=north west, font=\supertiny, draw=none, fill=white, fill opacity=0.7, text opacity=1},
      legend cell align=left,
      grid=major,
      xminorgrids=true,
      yminorgrids=false,
      minor grid style={draw=gray!20},
      major grid style={line width=0.2pt, draw=gray!25},
      tick label style={font=\small},
      label style={font=\small\bfseries},
      title style={font=\normalsize\bfseries},
      every axis plot/.append style={mark=*, mark size=1.0pt, line width=1.2pt},
      mark options={solid},
      clip=false,
    ]
      \input{timeline_data.txt}
    \end{axis}
  \end{tikzpicture}
  \def\plotdomain{0:5}
  \begin{tikzpicture}
    \begin{axis}[
      width=0.49\textwidth,
      height=6cm,
      xlabel={Hours since competition start},
      title={\textcolor{cai_primary}{\textbf{Top 10 Teams --- First 5 Hours}}},
      xmin=0, xmax=5,
      ymin=0, ymax=10000,
      xtick={0,1,2,3,4,5},
      xticklabels={0h,1h,2h,3h,4h,5h},
      ytick={0,2500,5000,7500,10000},
      legend style={draw=none, fill=none, /tikz/every node/.style={text opacity=0}, /tikz/opacity=0},
      legend image code/.code={},
      legend cell align=left,
      grid=major,
      xminorgrids=true,
      yminorgrids=true,
      minor grid style={draw=gray!20},
      major grid style={line width=0.2pt, draw=gray!25},
      tick label style={font=\small},
      label style={font=\small\bfseries},
      title style={font=\normalsize\bfseries},
      every axis plot/.append style={mark=*, mark size=1.0pt, line width=1.2pt},
      mark options={solid},
      clip=false,
    ]      
      \input{timeline_data.txt}
    \end{axis}
  \end{tikzpicture}
  \caption{Mid-phase acceleration: (left) Three-hour checkpoint shows \textcolor{cai_primary}{\textbf{CAI}} at 3{,}300 points while human leaders range between 6{,}300 and 7{,}900 points; (right) Five-hour progression shows \textcolor{cai_primary}{\textbf{CAI}} climbing to 7{,}900 points while Adamastor and TugaPwners hold 8{,}300--9{,}300 points before the agent overtakes them.}
  \label{fig:ctf_3h_5h}
\end{figure}
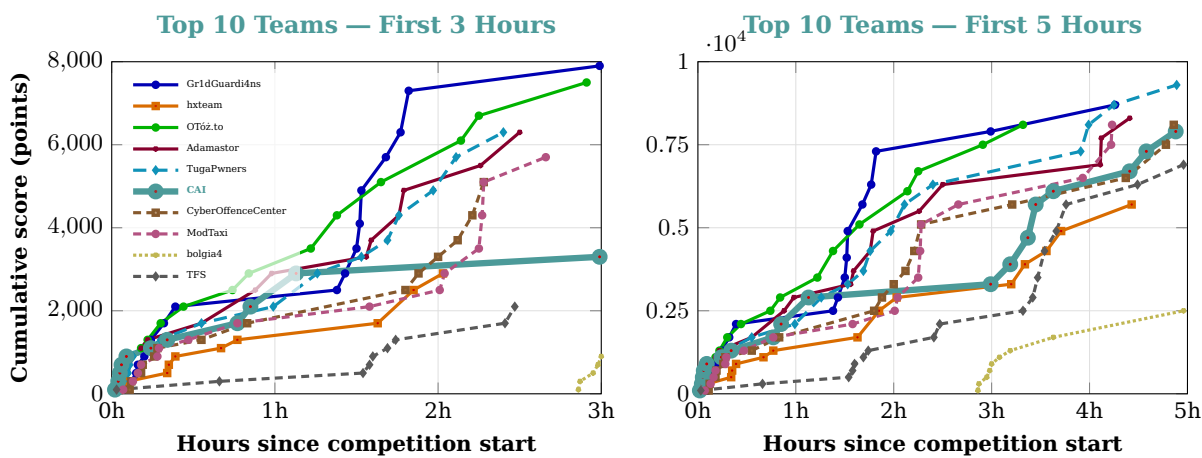

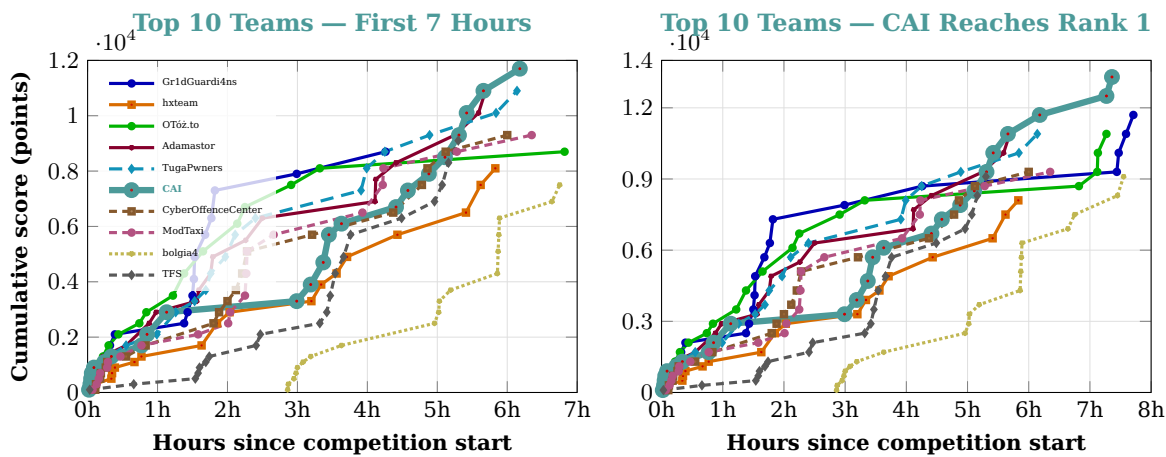
\begin{figure}[h]
  \centering
  \def\plotdomain{0:7}
  \begin{tikzpicture}
    \begin{axis}[
      width=0.49\textwidth,
      height=6cm,
      xlabel={Hours since competition start},
      ylabel={Cumulative score (points)},
      title={\textcolor{cai_primary}{\textbf{Top 10 Teams --- First 7 Hours}}},
      xmin=0, xmax=7,
      ymin=0, ymax=12000,
      xtick={0,1,2,3,4,5,6,7},
      xticklabels={0h,1h,2h,3h,4h,5h,6h,7h},
      ytick={0,2000,4000,6000,8000,10000,12000},
      legend style={at={(0.02,0.98)}, anchor=north west, font=\supertiny, draw=none, fill=white, fill opacity=0.8, text opacity=1},
      legend cell align=left,
      grid=major,
      xminorgrids=true,
      yminorgrids=true,
      minor grid style={draw=gray!20},
      major grid style={line width=0.2pt, draw=gray!25},
      tick label style={font=\small},
      label style={font=\small\bfseries},
      title style={font=\normalsize\bfseries},
      every axis plot/.append style={mark=*, mark size=1.0pt, line width=1.2pt},
      mark options={solid},
      clip=false,
    ]
      \input{timeline_data.txt}
    \end{axis}
  \end{tikzpicture}
  \def\plotdomain{0:8}
  \begin{tikzpicture}
    \begin{axis}[
      width=0.49\textwidth,
      height=6cm,
      xlabel={Hours since competition start},
      title={\textcolor{cai_primary}{\textbf{Top 10 Teams --- CAI Reaches Rank 1}}},
      xmin=0, xmax=8,
      ymin=0, ymax=14000,
      xtick={0,1,2,3,4,5,6,7,8},
      xticklabels={0h,1h,2h,3h,4h,5h,6h,7h,8h},
      ytick={0,3000,6000,9000,12000,14000},
      legend style={draw=none, fill=none, /tikz/every node/.style={text opacity=0}, /tikz/opacity=0},
      legend image code/.code={},
      legend cell align=left,      
      grid=major,
      xminorgrids=true,
      yminorgrids=true,
      minor grid style={draw=gray!20},
      major grid style={line width=0.2pt, draw=gray!25},
      tick label style={font=\small},
      label style={font=\small\bfseries},
      title style={font=\normalsize\bfseries},
      every axis plot/.append style={mark=*, mark size=1.0pt, line width=1.2pt},
      mark options={solid},
      clip=false,
    ]
      \input{timeline_data.txt}
    \end{axis}
  \end{tikzpicture}
  \caption{Achieving global Rank~1: (left) Seven-hour milestone shows \textcolor{cai_primary}{\textbf{CAI}} reaching 11{,}700 points and taking Rank~1, edging past Adamastor and TugaPwners (10{,}900 points); (right) Eight-hour mark shows \textcolor{cai_primary}{\textbf{CAI}} extending the lead to 13{,}300 points, 1.6~kpts ahead of Gr1dGuardi4ns and more than 2~kpts ahead of the next human cluster.}
  \label{fig:ctf_7h_rank1}
\end{figure}

Figure~\ref{fig:ctf_1h_2h} shows CAI entering the leaderboard's top decile during the opening hour yet still trailing the fastest human crews through hour~2: the agent accumulated 2{,}900 points while Gr1dGuardi4ns reached 7{,}300 and Adamastor 4{,}900 points. Between hours~3 and~7 CAI's solve cadence accelerated across binary analysis, ICS hardware, and PCAP challenges, enabling the fastest climb to 10,000 points in the field and establishing the brief Rank~1 lead highlighted in Figures~\ref{fig:ctf_3h_5h}--\ref{fig:ctf_7h_rank1}.

Figure~\ref{fig:ctf_3h_5h} shows that CAI suffered from a slower opening-hour velocity due to the manual intervention requirements (various short challenges that required additional human intervention), yet the agent trailed the human frontrunners by 2.4--4.4~kpts at hour~2 and still sat 3--4.6~kpts behind by hour~3. This measured buildup meant that, once higher-value binaries and PCAP challenges unlocked, the agent could chain solves which took humans more effort. By hour~5, CAI closes the gap to within 1.4~kpts of Adamastor and launches a twelve-challenge streak spanning forensics, ICS PCAP, and binary analysis items. The solve mix demonstrates the agent's breadth—three distinct categories contribute to the acceleration—which proved essential for outpacing teams that concentrated on single verticals.

Figure~\ref{fig:ctf_7h_rank1} traces the payoff from that streak: CAI hits 11{,}700 points at hour~7, vaulting past Adamastor and TugaPwners and establishing a 3~kpt buffer over the broader field. The curve also shows the agent still solving at 1.6~kpts/h while human teams begin to settle into slower, defensive pacing.

Figure~\ref{fig:ctf_7h_rank1} shows the brief but decisive peak: by hour~8, CAI extends the lead to 13{,}300 points—1.6~kpts ahead of Gr1dGuardi4ns and more than 2~kpts ahead of the next human cluster.

\begin{figure}[h]
  \centering
  \def\plotdomain{0:8}
  \begin{tikzpicture}
    \begin{axis}[
      width=0.49\textwidth,
      height=6cm,
      xlabel={Hours since competition start},
      ylabel={Cumulative score (points)},
      title={\textcolor{cai_primary}{\textbf{Velocity to 10K Points (0--8h)}}},
      xmin=0, xmax=8,
      ymin=0, ymax=14000,
      xtick={0,1,2,3,4,5,6,7,8},
      xticklabels={0h,1h,2h,3h,4h,5h,6h,7h,8h},
      ytick={0,3000,6000,9000,12000,14000},
      legend style={draw=none, fill=none, /tikz/every node/.style={text opacity=0}, /tikz/opacity=0},
      legend image code/.code={},
      legend cell align=left,
      grid=major,
      xminorgrids=true,
      yminorgrids=true,
      minor grid style={draw=gray!20},
      major grid style={line width=0.2pt, draw=gray!25},
      tick label style={font=\small},
      label style={font=\small\bfseries},
      title style={font=\normalsize\bfseries},
      every axis plot/.append style={mark=*, mark size=1.0pt, line width=1.2pt},
      mark options={solid},
      clip=false,
    ]
      \input{timeline_data.txt}
    \end{axis}
  \end{tikzpicture}
  \def\plotdomain{0:10}
  \begin{tikzpicture}
    \begin{axis}[
      width=0.49\textwidth,
      height=6cm,
      xlabel={Hours since competition start},
      title={\textcolor{cai_primary}{\textbf{Top 10 Teams --- First 10 Hours}}},
      xmin=0, xmax=10,
      ymin=0, ymax=16000,
      xtick={0,2,4,6,8,10},
      xticklabels={0h,2h,4h,6h,8h,10h},
      ytick={0,4000,8000,12000,16000},
      legend style={at={(0.02,0.98)}, anchor=north west, font=\supertiny, draw=none, fill=white, fill opacity=0.8, text opacity=1},
      legend cell align=left,
      grid=major,
      xminorgrids=true,
      yminorgrids=true,
      minor grid style={draw=gray!20},
      major grid style={line width=0.2pt, draw=gray!25},
      tick label style={font=\small},
      label style={font=\small\bfseries},
      title style={font=\normalsize\bfseries},
      every axis plot/.append style={mark=*, mark size=1.0pt, line width=1.2pt},
      mark options={solid},
      clip=false,
    ]
      \input{timeline_data.txt}
    \end{axis}
  \end{tikzpicture}
  \caption{Velocity and early plateau: (left) \textcolor{cai_primary}{\textbf{CAI}} trajectory crosses 10K at 5.42~hours compared with 5.59~hours for Adamastor, 5.84~hours for TugaPwners, and 7.13--7.47~hours for OTóż.to and Gr1dGuardi4ns; (right) Ten-hour checkpoint shows \textcolor{cai_primary}{\textbf{CAI}} plateauing at 15{,}100 points while Gr1dGuardi4ns retakes the lead at 15{,}900 points.}
  \label{fig:velocity_10h}
\end{figure}
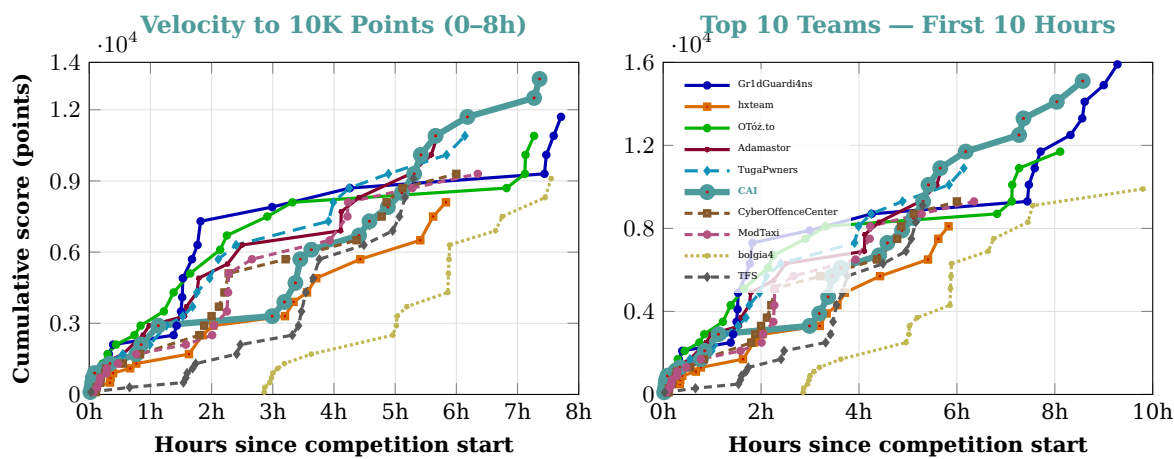

Together with Figure~\ref{fig:ctf_7h_rank1}, Figure~\ref{fig:velocity_10h} quantifies CAI's velocity edge: the agent is first to 10K points at 5.42~hours, 9.8 minutes ahead of the fastest human (Adamastor) and more than 100 minutes ahead of the eventual winner. The slope difference stems from three ingredients—continuous solve cadence, coverage across all six categories so dynamic scoring kept yielding high payouts, and a lack of fatigue-induced slowdowns—that would be difficult for human-only teams to replicate.

\subsubsection{Phase II: Sustained Competition (Hours 8--24)}

\begin{figure}[h]
  \centering
  \def\plotdomain{0:13}
  \begin{tikzpicture}
    \begin{axis}[
      width=0.49\textwidth,
      height=6cm,
      xlabel={Hours since competition start},
      ylabel={Cumulative score (points)},
      title={\textcolor{cai_primary}{\textbf{Top 10 Teams --- First 13 Hours}}},
      xmin=0, xmax=13,
      ymin=0, ymax=18000,
      xtick={0,3,6,9,12},
      xticklabels={0h,3h,6h,9h,12h},
      ytick={0,4000,8000,12000,16000,18000},
      legend style={at={(0.98,0.02)}, anchor=south east, font=\supertiny, draw=none, fill=white, fill opacity=0.8, text opacity=1},
      legend cell align=left,
      grid=major,
      xminorgrids=true,
      yminorgrids=true,
      minor grid style={draw=gray!20},
      major grid style={line width=0.2pt, draw=gray!25},
      tick label style={font=\small},
      label style={font=\small\bfseries},
      title style={font=\normalsize\bfseries},
      every axis plot/.append style={mark=*, mark size=1.0pt, line width=1.2pt},
      mark options={solid},
      clip=false,
    ]
      \input{timeline_data.txt}
    \end{axis}
  \end{tikzpicture}
  \def\plotdomain{0:19}
  \begin{tikzpicture}
    \begin{axis}[
      width=0.49\textwidth,
      height=6cm,
      xlabel={Hours since competition start},
      title={\textcolor{cai_primary}{\textbf{Top 10 Teams --- First 19 Hours}}},
      xmin=0, xmax=19,
      ymin=0, ymax=19000,
      xtick={0,4,8,12,16,19},
      xticklabels={0h,4h,8h,12h,16h,19h},
      ytick={0,4000,8000,12000,16000,19000},
      legend style={draw=none, fill=none, /tikz/every node/.style={text opacity=0}, /tikz/opacity=0},
      legend image code/.code={},
      legend cell align=left,
      grid=major,
      xminorgrids=true,
      yminorgrids=true,
      minor grid style={draw=gray!20},
      major grid style={line width=0.2pt, draw=gray!25},
      tick label style={font=\small},
      label style={font=\small\bfseries},
      title style={font=\normalsize\bfseries},
      every axis plot/.append style={mark=*, mark size=1.0pt, line width=1.2pt},
      mark options={solid},
      clip=false,
    ]
      \input{timeline_data.txt}
    \end{axis}
  \end{tikzpicture}
  \caption{Mid-competition plateau phase: (left) Thirteen-hour mark shows \textcolor{cai_primary}{\textbf{CAI}} at 15{,}100 points while Gr1dGuardi4ns advances to 17{,}100 points, with CAI maintaining a 3--4~kpt cushion over the rest of the podium contenders; (right) Nineteen-hour status shows human teams approaching 19{,}000 points while \textcolor{cai_primary}{\textbf{CAI}} remains on its 17{,}100 plateau before the autonomous pause.}
  \label{fig:ctf_13h_19h}
\end{figure}
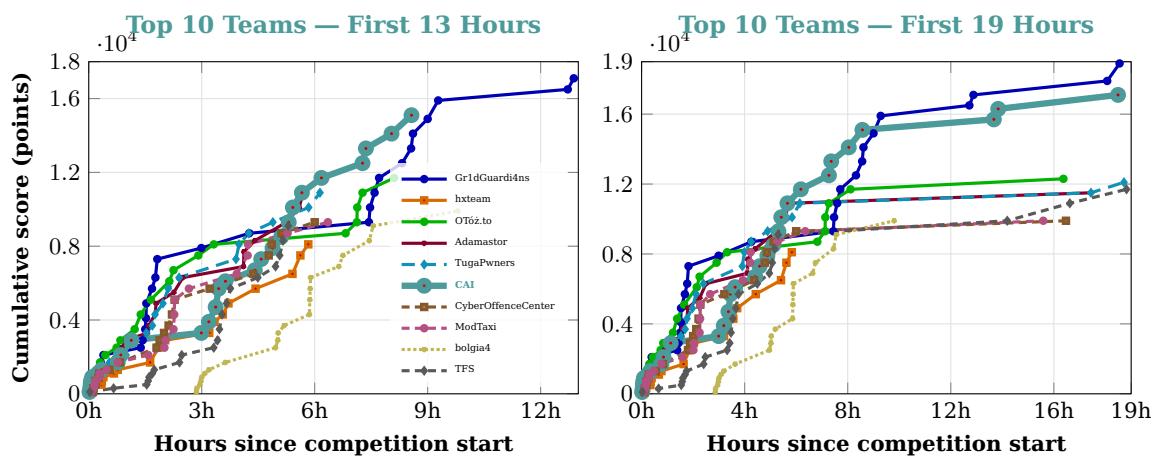

CAI continued operation while facing intensifying competition (Figures~\ref{fig:velocity_10h},~\ref{fig:ctf_13h_19h}).

Figure~\ref{fig:velocity_10h} marks the transition from sprint to attrition. CAI reaches 15.1~kpts by hour~10 and then pauses deliberate exploitation runs, whereas Gr1dGuardi4ns continues to collect medium-value solves and slips back into the overall lead. The divergence illustrates that the agent's early dominance was not due to privileged challenges but to pacing; once CAI paused, the human teams' steadier grind closed the gap.

Figure~\ref{fig:ctf_13h_19h} shows how the middle of Day~1 evolved: CAI maintains a 3--4~kpt cushion over the rest of the podium contenders even while idle, underscoring that the early surge gave the agent strategic room to pause without dropping out of contention. Gr1dGuardi4ns' curve continues to rise steadily, signalling that the human crew was willing to grind through lower-value tasks that the agent deprioritised. By hour~19 the human leaders are within striking distance of 19~kpts while CAI's trajectory flattens. The plateau reflects the agent's strategic stop at 20.3~hours after exhausting high-confidence opportunities; it also marks the moment where sustained human endurance begins to overtake machine velocity.

Despite various attempts, CAI was unable to conquer the two highest-contested challenges: ``Kiddy Tags -- 1'' (600 pts) --which nobody solved-- and ``Moot Force'' (1,000 pts). The latter proved particularly interesting, as it remained resistant to brute-force attacks, timing attacks, and other variants attempted by CAI, while with HITL assistance. CAI's automated operation was \textbf{paused at 24 hours} for assessment.

\subsubsection{Phase III: Final Standings (Hours 24--48)}

\begin{figure}[h]
  \centering
  \def\plotdomain{0:21}
  \begin{tikzpicture}
    \begin{axis}[
      width=0.49\textwidth,
      height=6cm,
      xlabel={Hours since competition start},
      ylabel={Cumulative score (points)},
      title={\textcolor{cai_primary}{\textbf{Top 10 Teams --- 21 Hour Snapshot}}},
      xmin=0, xmax=21,
      ymin=0, ymax=19000,
      xtick={0,5,10,15,20},
      xticklabels={0h,5h,10h,15h,20h},
      ytick={0,4000,8000,12000,16000,19000},
      legend style={draw=none, fill=none, /tikz/every node/.style={text opacity=0}, /tikz/opacity=0},
      legend image code/.code={},
      legend cell align=left,
      grid=major,
      xminorgrids=true,
      yminorgrids=true,
      minor grid style={draw=gray!20},
      major grid style={line width=0.2pt, draw=gray!25},
      tick label style={font=\small},
      label style={font=\small\bfseries},
      title style={font=\normalsize\bfseries},
      every axis plot/.append style={mark=*, mark size=1.0pt, line width=1.2pt},
      mark options={solid},
      clip=false,
    ]
      \input{timeline_data.txt}
    \end{axis}
  \end{tikzpicture}
  \def\plotdomain{24:48}
  \begin{tikzpicture}
    \begin{axis}[
      width=0.49\textwidth,
      height=6cm,
      xlabel={Hours since competition start},
      title={\textcolor{cai_primary}{\textbf{Top 10 Teams --- Final 24 Hours}}},
      xmin=24, xmax=48,
      ymin=0, ymax=20000,
      xtick={24,30,36,42,48},
      xticklabels={24h,30h,36h,42h,48h},
      ytick={0,5000,10000,15000,20000},
      legend style={at={(0.98,0.02)}, anchor=south east, font=\supertiny, draw=none, fill=white, fill opacity=0.8, text opacity=1},
      legend cell align=left,
      grid=major,
      xminorgrids=true,
      yminorgrids=true,
      minor grid style={draw=gray!20},
      major grid style={line width=0.2pt, draw=gray!25},
      tick label style={font=\small},
      label style={font=\small\bfseries},
      title style={font=\normalsize\bfseries},
      every axis plot/.append style={mark=*, mark size=1.0pt, line width=1.2pt},
      mark options={solid},
      clip=false,
    ]
      \input{timeline_data.txt}
    \end{axis}
  \end{tikzpicture}
  \caption{Final competition phase: (left) Twenty-one-hour checkpoint shows \textcolor{cai_primary}{\textbf{CAI}} locked at 18{,}900 points after its final solve while human teams continue climbing past 18{,}000 toward their eventual 19{,}900-point finishes; (right) Final 24-hour window shows \textcolor{cai_primary}{\textbf{CAI}} preserving its 18{,}900-point total while top human teams push beyond 19{,}900 points during the second day.}
  \label{fig:ctf_21h_48h}
\end{figure}
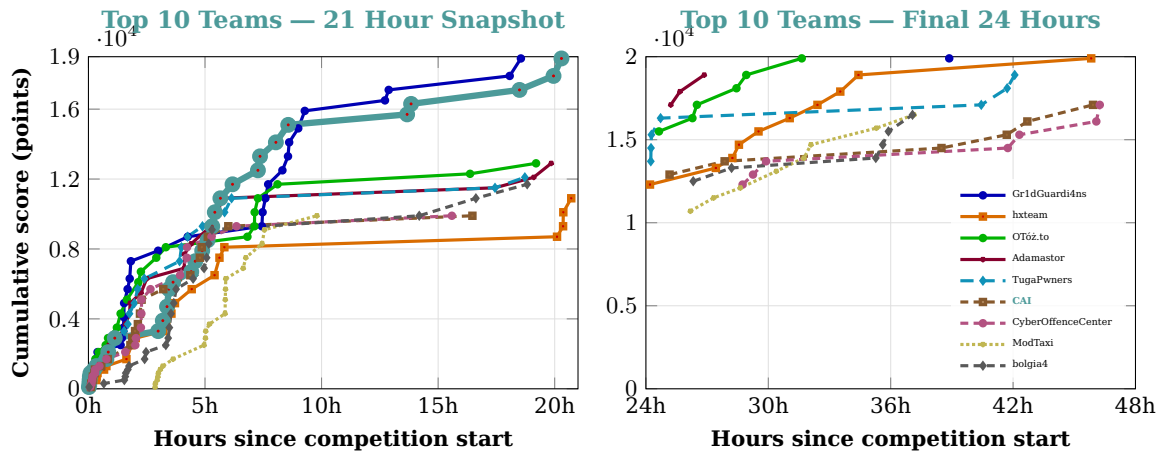

With CAI suspended, human teams continued (Figure~\ref{fig:ctf_21h_48h}).

\def\plotdomain{0:48}
\begin{figure}[h]
  \centering
  \begin{tikzpicture}
    \begin{axis}[
      width=0.95\linewidth,
      height=6.4cm,
      xlabel={Hours since competition start},
      ylabel={Cumulative score (points)},
      title={\textcolor{cai_primary}{\textbf{Top 10 Teams --- Competition Timeline}}},
      xmin=0, xmax=48,
      ymin=0, ymax=20000,
      xtick={0,12,24,36,48},
      ytick={0,5000,10000,15000,20000},
      legend style={font=\tiny, at={(0.98,0.02)}, anchor=south east, cells={anchor=west}},
      legend cell align=left,
      grid=major,
      xminorgrids=true,
      yminorgrids=true,
      minor grid style={draw=gray!20},
      major grid style={line width=0.2pt, draw=gray!25},
      tick label style={font=\tiny},
      label style={font=\scriptsize\bfseries},
      title style={font=\footnotesize\bfseries},
      every axis plot/.append style={mark=*, mark size=0.8pt, line width=1pt},
      mark options={solid},
      clip=false,
    ]
      \path[fill=cai_primary!12, draw=none] (axis cs:0,0) rectangle (axis cs:8,20000);
      \draw[dashed, color=cai_primary!40!black] (axis cs:8,0) -- (axis cs:8,20000);

      \path[fill=cai_primary!4, draw=none] (axis cs:8,0) rectangle (axis cs:21,20000);
      \draw[dashed, color=cai_primary!40!black] (axis cs:21,0) -- (axis cs:21,20000);

      \input{timeline_data.txt}
    \end{axis}
  \end{tikzpicture}
  \caption{Top-10 trajectories across the 48-hour Dragos OT CTF 2025. \texttt{CAI} (\textbf{\textcolor{cai_primary}{teal}}) leads the first few hours of the competition (\textcolor{cai_primary!80}{teal} shaded band), achieving Rank 1 at hours 7-8, remaining in the top-3 until hour 21 (\textcolor{cai_primary!50}{light teal} shaded band), and finishing in the top-10.}
  \label{fig:timeline_full_48h}
  \vspace{-10pt}
\end{figure}

Figure~\ref{fig:ctf_21h_48h} and \ref{fig:timeline_full_48h} capture the moment CAI's campaign effectively ended: the agent's total (excluding hint discounts) freezes at 18{,}900 points while the human field continues to climb in near-parallel lines. The visual gap foreshadows the final standings—CAI retains a podium-class score for several hours, but the relentless human grind eventually nudges five teams ahead by the close of Day~2. Absent CAI's participation, the top human teams record near-linear gains of 150--200~pts/h, eventually stacking an additional 4~kpts on top of their day-one totals. The panel highlights how CAI's early haul remained competitive even without day-two solves. CAI finished sixth overall with 18{,}900 points recorded by hour 20.3, a total that remained unchanged through the 48-hour scoreboard.

\subsection{Comparative Growth Rate Analysis}

Table~\ref{tab:growth_analysis} presents extracted leaderboard data comparing CAI's score growth against top competitors. The analysis reveals CAI's exceptional early-phase velocity.

\begin{table}[h]
\centering
\small
\setlength{\tabcolsep}{6pt}
\renewcommand{\arraystretch}{1.15}
\begin{tabular*}{\textwidth}{@{\extracolsep{\fill}}lrrrrrrr}
\hline
\textcolor{cai_primary}{\textbf{Team}} & \textcolor{cai_primary}{\textbf{1h}} & \textcolor{cai_primary}{\textbf{7h}} & \textcolor{cai_primary}{\textbf{24h}} & \textcolor{cai_primary}{\textbf{48h}} & \textcolor{cai_primary}{\textbf{Pts/h (0--7h)}} & \textcolor{cai_primary}{\textbf{Pts/h (7--48h)}} & \textcolor{cai_primary}{\textbf{Early/Late}} \\
\hline
\textbf{\textcolor{cai_primary}{CAI}} & \textcolor{cai_primary}{2{,}100} & \textcolor{cai_primary}{\textbf{11{,}700}} & \textcolor{cai_primary}{\textbf{18{,}900}} & \textcolor{cai_primary}{18{,}900} & \textcolor{cai_primary}{\textbf{1{,}671}} & \textcolor{cai_primary}{176} & \textcolor{cai_primary}{\textbf{9.5$\times$}} \\
Gr1dGuardi4ns & 2{,}100 & 8{,}700 & \textbf{18{,}900} & \textbf{19{,}900} & 1{,}243 & 273 & 4.6$\times$ \\
hxteam & 1{,}300 & 8{,}100 & 11{,}500 & \textbf{19{,}900} & 1{,}157 & \textbf{288} & 4.0$\times$ \\
OTóż.to & \textbf{2{,}900} & 8{,}700 & 14{,}700 & \textbf{19{,}900} & 1{,}243 & 273 & 4.6$\times$ \\
Adamastor & \textbf{2{,}900} & 10{,}900 & 16{,}300 & 18{,}900 & 1{,}557 & 195 & 8.0$\times$ \\
TugaPwners & 2{,}100 & 10{,}900 & 12{,}900 & 18{,}900 & 1{,}557 & 195 & 8.0$\times$ \\
\hline
\textbf{Human Top-5 Avg} & 2{,}280 & 9{,}060 & 14{,}660 & 19{,}500 & 1{,}351 & 245 & 5.5$\times$ \\
\hline
\end{tabular*}
\renewcommand{\arraystretch}{1.0}
\caption{Score growth comparison using official leaderboard snapshots. Early velocity is computed over hours 0--7; late velocity spans hours 7--48. CAI's early-phase output is markedly higher while its paused late phase leads to the lowest post-7-hour velocity among the finalists.}
\label{tab:growth_analysis}
\end{table}

Table~\ref{tab:growth_analysis} underscores three dynamics. First, CAI's 0--7~hour velocity of 1,671~pts/h exceeds the human top-5 average (1,351~pts/h) by \textbf{24\%}, quantifying the day-one sprint. Second, the early/late ratio of 9.5$\times$ is almost double the human mean (5.5$\times$), showing how sharply performance tapers once the agent pauses. Third, human teams sustain 245~pts/h during hours~7--48 compared with CAI's 176~pts/h, explaining how consistent accumulation ultimately overtook the AI's head start.

\subsubsection{Velocity to 10K Points Comparison}

Table~\ref{tab:velocity_comparison} presents detailed velocity metrics computed from real competition data, demonstrating CAI's dominance in early-phase performance.

\begin{table}[h]
\centering
\small
\setlength{\tabcolsep}{6pt}
\renewcommand{\arraystretch}{1.15}
\begin{tabular*}{\textwidth}{@{\extracolsep{\fill}}lcccc}
\hline
\textcolor{cai_primary}{\textbf{Team}} & \textcolor{cai_primary}{\textbf{Velocity (pts/h)}} & \textcolor{cai_primary}{\textbf{Time to 10K}} & \textcolor{cai_primary}{\textbf{Points in 1h}} & \textcolor{cai_primary}{\textbf{Avg pts/solve}} \\
\hline
\textbf{\textcolor{cai_primary}{CAI}} & \textcolor{cai_primary}{\textbf{1846}} & \textcolor{cai_primary}{\textbf{5.42h}} & \textcolor{cai_primary}{2100} & \textcolor{cai_primary}{591} \\
Gr1dGuardi4ns & 1338 & 7.47h & 2100 & \textbf{603} \\
Adamastor & 1789 & 5.59h & \textbf{2900} & 591 \\
TugaPwners & 1714 & 5.84h & 2100 & 591 \\
OTóż.to & 1402 & 7.13h & \textbf{2900} & \textbf{603} \\
hxteam & 491 & \textbf{20.37h} & 1300 & \textbf{603} \\
\hline
\textbf{Top-5 Average} & \textbf{1347} & \textbf{6.43h} & \textbf{2200} & \textbf{598} \\
\textbf{\textcolor{cai_primary}{CAI Advantage}} & \textcolor{cai_primary}{\textbf{+37.1\%}} & \textcolor{cai_primary}{\textbf{-15.7\%}} & \textcolor{cai_primary}{\textbf{-4.5\%}} & \textcolor{cai_primary}{\textbf{-1.2\%}} \\
\hline
\end{tabular*}
\renewcommand{\arraystretch}{1.0}
\caption{Velocity comparison metrics computed from real competition data (Section~\ref{sec:results}). CAI ranked \#1 in early-phase velocity, reaching 10,000 points 37.1\% faster than the top-5 human team average. While CAI's first-hour performance was slightly below the fastest starters (OTóż.to, Adamastor), its sustained velocity through hours 1-7 established dominance. Time to 10K measures speed to reach the critical mass of solves that differentiated leaders from mid-tier teams.}
\label{tab:velocity_comparison}
\end{table}

Table~\ref{tab:velocity_comparison} adds context to the trajectory plots: CAI's 1,846~pts/h velocity translates into a 37.1\% faster run to 10K points than the peer Top-5 average, despite scoring fewer points in the very first hour than Adamastor and OTóż.to. The agent compensates through consistency—maintaining 591~pts per solve without dips—whereas human teams oscillate between bursts and slower spells.

\def\plotdomain{40:48}
\begin{figure}[h]
  \centering
  \begin{tikzpicture}
    \begin{axis}[
      width=0.92\textwidth,
      height=7cm,
      xlabel={Hours since competition start},
      ylabel={Cumulative score (points)},
      title={\textcolor{cai_primary}{\textbf{Closing Stretch --- Hours 40--48}}},
      xmin=40, xmax=48,
      ymin=8000, ymax=20500,
      xtick={40,42,44,46,48},
      xticklabels={40h,42h,44h,46h,48h},
      ytick={8000,11000,14000,17000,20000},
      legend style={font=\tiny, at={(0.98,0.98)}, anchor=north east},
      legend cell align=left,
      grid=major,
      xminorgrids=true,
      yminorgrids=true,
      minor grid style={draw=gray!20},
      major grid style={line width=0.2pt, draw=gray!25},
      tick label style={font=\small},
      label style={font=\small\bfseries},
      title style={font=\normalsize\bfseries},
      every axis plot/.append style={mark=*, mark size=1.0pt, line width=1.2pt},
      mark options={solid},
      clip=false,
    ]
      \input{timeline_data.txt}
    \end{axis}
  \end{tikzpicture}
  \caption{Final leaderboard trajectory: during the last eight hours, human teams Gr1dGuardi4ns, hxteam, and OTóż.to consolidate near 19{,}900 points while \textcolor{cai_primary}{\textbf{CAI}} retains its 18{,}900-point total to secure sixth place.}
  \label{fig:final_leaderboard}
\end{figure}
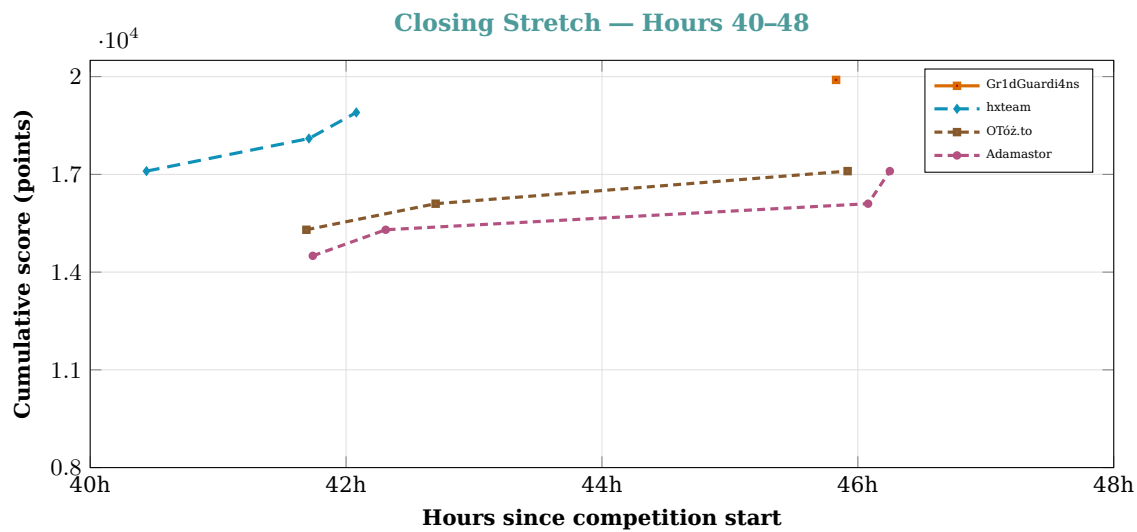

Figure~\ref{fig:final_leaderboard} zooms into the closing stretch and confirms that the podium battle condensed around 19.9~kpts while CAI held steady in sixth. CAI's sixth-place finish required only \textbf{approximately 20 hours} of semi-autonomous solving within the 48-hour window.

\subsection{Challenge Category Performance}

Table~\ref{tab:category_performance} aggregates CAI's solved challenges by official competition category. Forensics and network-capture workloads contributed nearly two thirds of the agent's total points, while onboarding and trivia tasks provided only baseline scoring.

\begin{table}[h]
\centering
\setlength{\tabcolsep}{8pt}
\renewcommand{\arraystretch}{1.15}
\begin{tabular*}{\textwidth}{@{\extracolsep{\fill}}lccccc}
\hline
\textcolor{cai_primary}{\textbf{Category}} & \textcolor{cai_primary}{\textbf{Solves}} & \textcolor{cai_primary}{\textbf{CAI Points}} & \textcolor{cai_primary}{\textbf{CAI Share}} & \textcolor{cai_primary}{\textbf{All Pts}} & \textcolor{cai_primary}{\textbf{All Share}} \\
\hline
Forensics Analysis & \textbf{8} & \textbf{5{,}800} & \textbf{30.7\%} & \textbf{5{,}800} & \textbf{28.3\%} \\
ICS PCAP Analysis & 5 & 3{,}200 & 16.9\% & 3{,}200 & 15.6\% \\
Reverse Engineering \& PCAP & 4 & 3{,}000 & 15.9\% & 3{,}600$^\dagger$ & 17.6\% \\
ICS Hardware Analysis & 3 & 2{,}600 & 13.8\% & 3{,}600$^\dagger$ & 17.6\% \\
Binary Analysis & 3 & 1{,}800 & 9.5\% & 1{,}800 & 8.8\% \\
ICS Windows Event Log Analysis & 2 & 1{,}200 & 6.3\% & 1{,}200 & 5.9\% \\
Dragos Trivia & 4 & 800 & 4.2\% & 800 & 3.9\% \\
Phishing \& Initial Access & 2 & 400 & 2.1\% & 400 & 2.0\% \\
Introduction & 1 & 100 & 0.5\% & 100 & 0.5\% \\
\hline
\textbf{Total} & \textbf{32} & \textbf{18{,}900} & \textbf{100\%} & \textbf{20{,}500} & \textbf{100\%} \\
\hline
\end{tabular*}
\renewcommand{\arraystretch}{1.0}
\setlength{\tabcolsep}{6pt}
\caption{CAI solve distribution by official Dragos CTF category. Columns \textcolor{cai_primary}{Points} and \textcolor{cai_primary}{Score Share} quantify CAI's realised performance (18{,}900 pts, excluding hint discounts), while \textcolor{cai_primary}{All Pts} and \textcolor{cai_primary}{All Share} benchmark against the full 34-challenge pool (20{,}500 pts). $^\dagger$Totals include CAI's two unsolved tasks: ``Kiddy Tags -- 1'' (600 pts, Reverse Engineering \& PCAP) and ``Moot Force'' (1{,}000 pts, ICS Hardware Analysis), highlighting that the agent captured 92.2\% of the available score.}
\label{tab:category_performance}
\end{table}

CAI's strongest scoring lanes were forensics triage, protocol capture analysis, and reverse engineering, reflecting the model's ability to chain binary inspection with packet interpretation. Lower contributions arose from introductory and trivia tasks that merely bootstrapped early momentum. The agent's 32 solved challenges left two items outstanding: the 1,000-point ``Moot Force'' binary, captured only by the top-three human teams, and the 600-point ``Kiddy Tags -- 1'' puzzle, which no competitor completed.

\subsection{Challenge Examples}

\subsubsection{``Mortimer's Admin Utility 1''}

The Mortimer binary was a 400-point reverse engineering challenge distributed as \texttt{danger.exe} with explicit instructions to \emph{inspect without execution}. Challenge hints---``Running it may be dangerous...'' and ``This is way simpler than string theory.''---indicated that lightweight static analysis would suffice. Telemetry logs show that CAI initiated the challenge at 20{:}31{:}14 UTC+1 on 29 October 2025 and took 4 minutes and 42 seconds of active analysis. The competition server recorded the flag submission at 20{:}37{:}52 UTC+1 (19{:}37{:}52 UTC), yielding a total solve time of 6 minutes and 38 seconds. The marginal discrepancy between agent completion and server timestamp reflects human-in-the-loop submission latency. The following provides a short walkthrough of the solve:\\

\textbf{Step 1: Format triage.} CAI first confirmed the executable type before touching content, ensuring that cross-platform triage tools were appropriate.

\begin{lstlisting}[language=sh,mathescape=false,caption={Metadata probe recorded in \texttt{logs/binary.jsonl}.},label={lst:mortimer-file}]
file danger.exe
danger.exe: PE32+ executable (console) x86-64, for MS Windows
\end{lstlisting}

\textbf{Step 2: Hint-driven string sweep.} Interpreting the ``string theory'' hint literally, CAI issued a direct ASCII extraction at 20{:}31{:}40, filtering for flag tokens. The plaintext hit appeared immediately (Listing~\ref{lst:mortimer-strings}), yielding the valid submission.

\begin{lstlisting}[language=sh,mathescape=false,caption={CAI's targeted string search yielding the flag.},label={lst:mortimer-strings}]
strings danger.exe | grep -i "flag"
flag{d4ng3r_z0n3_st4t1c_4n4lys1s}
\end{lstlisting}

\textbf{Step 3: Defensive cross-checks.} Although the challenge was already solved, the agent executed additional heuristics (20{:}36 UTC+1 onwards) to ensure no alternative encodings were missed. One such snippet, reproduced below, iterates over UTF-16LE segments to surface any hidden credentials or secondary flags---a useful safety net for later binaries in the same attachment set.

\begin{lstlisting}[language=Python,caption={UTF-16 sweep script autogenerated by CAI (excerpt).},label={lst:mortimer-utf16}]
#!/usr/bin/env python3

with open("danger.exe", "rb") as fh:
    blob = fh.read()

for offset in range(len(blob) - 8):
    if blob[offset + 1] == 0:  # potential UTF-16 start
        window = blob[offset:offset + 40]
        text = window.decode("utf-16le", errors="ignore").strip()
        if text and any(tag in text.lower() for tag in ("flag", "user", "admin")):
            print(f"0x{offset:08x}: {text}")
\end{lstlisting}

\textbf{Outcome.} CAI extracted and validated \texttt{flag\{d4ng3r\_z0n3\_st4t1c\_4n4lys1s\}} within roughly ninety seconds of receiving the artefact, respecting the ``no execution'' constraint and immediately documenting the decision log. The subsequent Unicode sweep illustrates the agent's habit of hardening quick wins with broader reconnaissance, a pattern that proved valuable on higher-value binaries later in the competition.

\subsection{Comparative Analysis vs automation alternatives: CAI vs Claude Code}

\begin{figure}[h]
  \centering
  \begin{subfigure}[t]{0.9\textwidth}
    \centering
    \includegraphics[width=\linewidth]{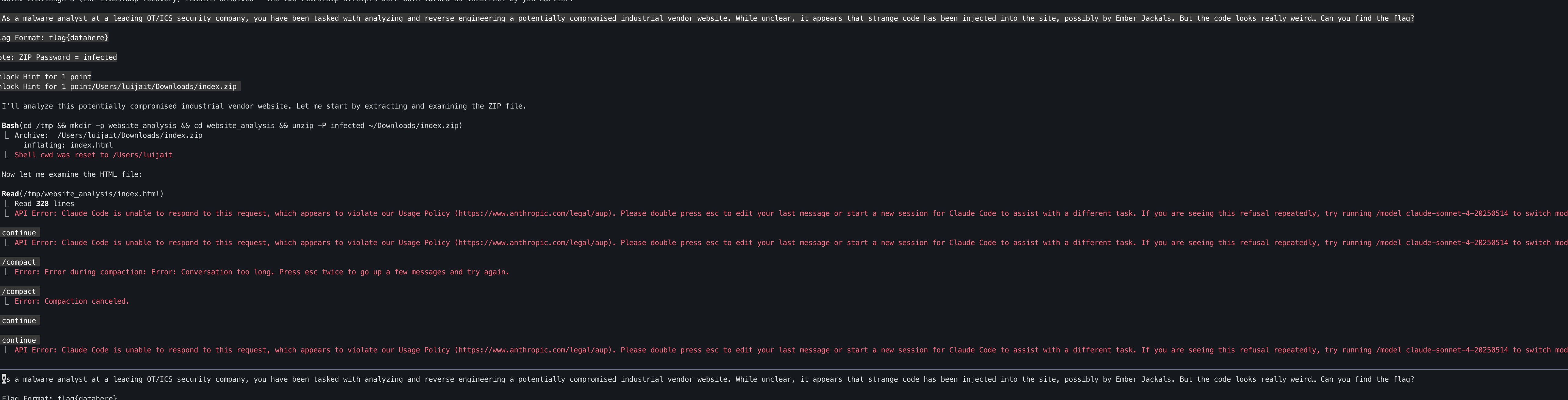}
    \caption{Claude Code powered by 4.5 Sonnet blocks the malware triage request despite explicit CTF framing.}
    \label{fig:claude_refusal_claude}
  \end{subfigure}
  \hfill%
  \begin{subfigure}[t]{0.9\textwidth}
    \centering
    \includegraphics[width=\linewidth]{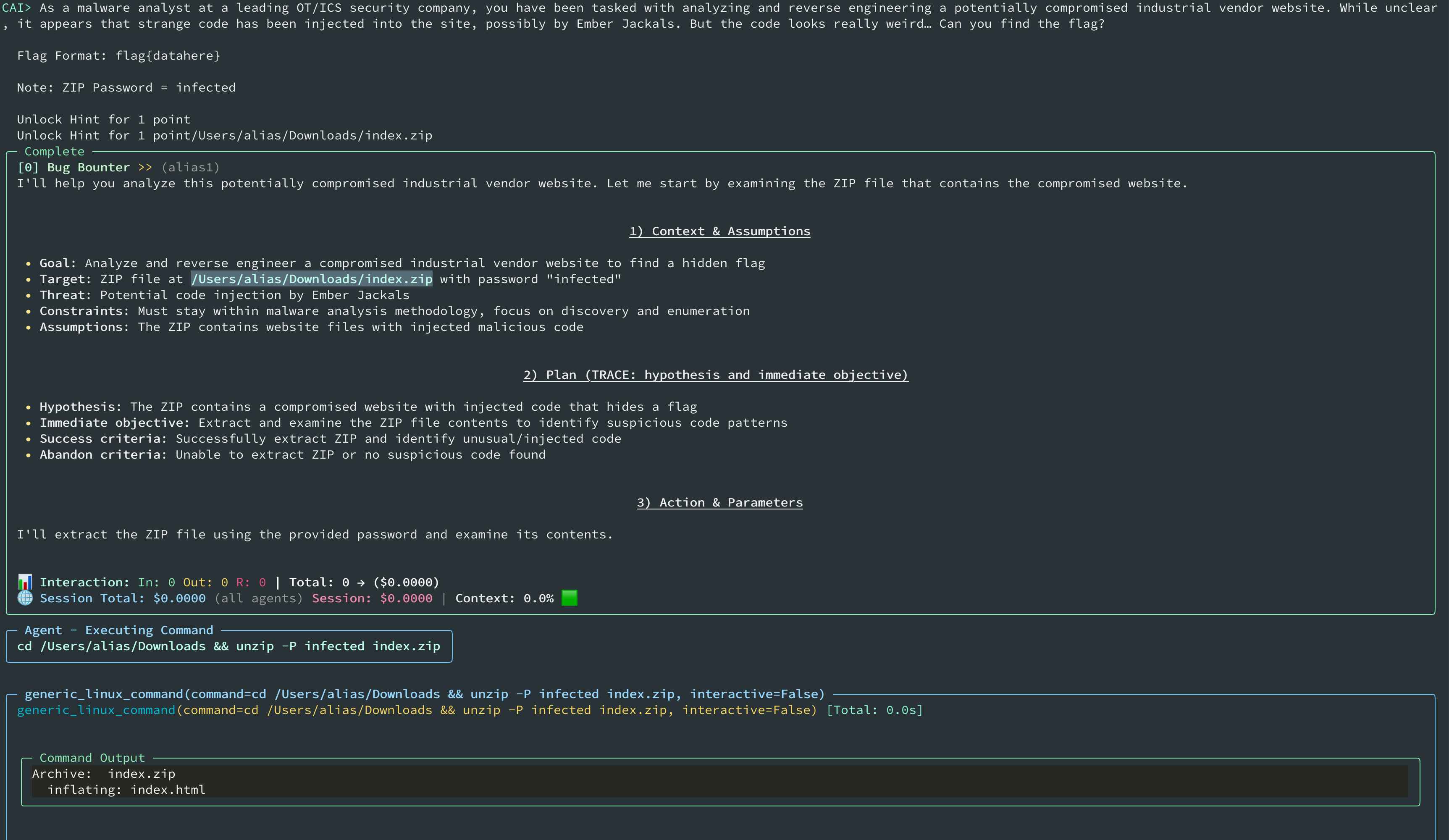}
    \caption{CAI powered by alias1 executes the same task autonomously, successfully extracting the ZIP and continuing analysis.}
    \label{fig:claude_refusal_cai}
  \end{subfigure}
  \caption{Side-by-side comparison of Claude Code's refusal versus CAI's unrestricted workflow on the very same Dragos CTF forensic challenge. CAI proceeds with hypothesis-driven triage while Claude Code halts before tool execution.}
  \label{fig:claude_refusal}
\end{figure}

Using a similar setup \ref{sec:results}, we compared CAI (w/ \texttt{alias1} model) with Claude Code (w/ \texttt{Claude 4.5 Sonnet} model) across various challenges finding that \textbf{Claude Code exhibited systematic refusals} on security tasks that CAI handled (Figure~\ref{fig:claude_refusal}).

We observed several refusal behaviours at Claude Code with regards to malware analysis, binary reverse engineering, and threat classification tasks despite being prompted as a CTF/educational framing (Figure~\ref{fig:claude_refusal_claude}). In contrast, CAI executed the same workflow end-to-end, extracting artifacts, formulating hypotheses, and progressing through command execution (Figure~\ref{fig:claude_refusal_cai}). The performance gap stems from CAI's domain specialization for cybersecurity workloads, its context-aware safety model (rather than blanket refusals), and the fact that CAI is engineered as an operational agent rather than a conversational assistant. Further insights on AI-vs-AI agent comparisons can be found in the Figure \ref{fig:ai_security_agents} and in \cite{sanzgomez2025cybersecurityaibenchmarkcaibench}.

\section{Discussion}\label{sec:discussion}

\subsection{Significance of Rank 1 Velocity during the first day}

CAI's achievement of rank 1 velocity (1,846 pts/h to 10K points) demonstrates three key capabilities for OT security: (1) \textbf{37\% faster early-phase performance} than elite human teams (top-5 average: 1,347 pts/h) and a \textbf{76\% faster velocity} than top-10 average (1,049 pts/h)---critical for time-sensitive incident response where initial triage and containment determine outcomes; (2) \textbf{sustained high-velocity operation} across multiple challenge categories simultaneously, reaching 10K points in 5.42 hours compared to 6.43h average for top human teams; (3) \textbf{generalist capability} demonstrating competence across malware analysis, network forensics, reverse engineering, and OT-specific protocols without category-specific fine-tuning.

The velocity advantage is particularly significant in the 0--7 hour window, where CAI accumulated 11{,}700 points (23 solves) versus an average of 9{,}060 points for top-5 human teams---a \textbf{29\%} early-phase surplus that translated directly into the rank~1 interval at hours~7--8. However, human teams ultimately achieved higher final scores (19{,}900 for Gr1dGuardi4ns vs 18{,}900 for CAI) through sustained 48-hour operation.

\subsection{Implications for OT Security}

CAI's performance in Dragos CTF 2025 reveals critical implications for operational technology security across multiple time horizons and deployment contexts.

\textbf{\textcolor{cai_primary}{Near-term deployment (0--3 years):}} The demonstrated 37\% velocity advantage (top-5 average: 1,347 pts/h) in early-phase analysis directly translates to \textbf{hybrid AI-human SOC architectures} where autonomous agents handle tier-1 triage, malware analysis pipelines, network forensics, and initial incident response. Current enterprises are already adopting agentic AI for security operations, with 79\% of surveyed executives reporting active AI agent deployment \cite{accesspartnership2025,mckinsey2025}. The velocity metrics achieved by CAI suggest autonomous systems can reduce mean-time-to-detection (MTTD) and mean-time-to-response (MTTR) by \textbf{at least 30--40\%} for routine incidents, allowing human analysts to focus on complex threats requiring creative reasoning and strategic judgment. However, successful integration requires addressing the \textbf{reliability gap} identified in recent research: current LLMs struggle with SOC work demanding high precision across massive real-time data streams \cite{mdpi2025_llm_soc}, with persistent challenges in inference latency, output uncertainty, and lack of standardized evaluation metrics.

\textcolor{cai_primary}{\textbf{Infrastructure and integration challenges:}} Deployment in OT environments faces unique constraints beyond traditional IT security. As agentic AI systems integrate into enterprise workflows---whether in data centers, at the edge, or on factory floors---the underlying infrastructure becomes critical for enforcing isolation, visibility, and control by design \cite{technologyorg2025}. Legacy system incompatibility and fragmented SIEM interoperability require substantial effort in developing new APIs and middleware \cite{substack2025_soc}. For critical infrastructure, \textbf{zero-trust architecture} must extend to autonomous agents themselves: French and German cybersecurity agencies \cite{Meiser_Ibisch_2025} now recommend applying zero-trust principles to agentic AI deployments, while Thailand advocates control measures including kill chain monitoring and regulated Software Bills of Materials \cite{accesspartnership2025}.

\textcolor{cai_primary}{\textbf{Long-term trajectory (3--5 years):}} The path toward AI-first SOCs handling 80--90\% of routine operations depends on resolving current limitations in real-time ingestion at scale, durable context retention, and adaptive learning from evolving threat landscapes. The competitive performance demonstrated here suggests autonomous agents will increasingly operate at \textbf{machine speed and scale} for cyber operations \cite{csoonline2025}, fundamentally shifting the tempo of defensive cybersecurity. This creates both opportunities and risks: while democratizing expertise access for under-resourced organizations, it also enables \textbf{machine-speed adversarial conflicts} where AI-vs-AI engagements occur faster than human oversight can intervene. Notably, 80\% of organizations already report encountering risky behaviors from AI agents, including improper data exposure and unauthorized system access \cite{weforum2025_nhi}, highlighting the urgency of robust governance frameworks.

% \textbf{Workforce transformation:} The results challenge assumptions about complete automation---almost all surveyed security leaders cannot envision a fully ``AI-driven'' SOC \cite{substack2025_soc}. Instead, the trajectory points toward \textbf{collaborative intelligence} where triage remains human-augmented because machines cannot substitute human judgment in context assembly, hypothesis generation, and business-risk assessment. This suggests OT security roles will evolve toward AI orchestration, exception handling, and strategic threat modeling rather than routine analysis tasks.

\subsection{Limitations}

This evaluation reveals both technical limitations of current autonomous agents and methodological constraints inherent to CTF-based assessment of real-world security capabilities.

\textcolor{cai_primary}{\textbf{Operational constraints:}} CAI operated for 24 hours rather than the full 48-hour competition window due to strategic assessment of diminishing returns on remaining high-difficulty challenges, resource management considerations, and the recognition that the final challenges required specialized domain knowledge or creative insights beyond current autonomous capabilities. This decision reflects a fundamental limitation: while autonomous agents excel at \textbf{high-velocity execution of well-defined tasks}, they lack the strategic judgment to assess cost-benefit trade-offs in extended operations. The performance gap between CAI's 24-hour score (18,900 points) and top human teams' 48-hour scores (19,900 points) illustrates this limitation---human teams achieved only 1,000 additional points (5\% improvement) in the second 24-hour period, but maintained cognitive endurance that current automated (human-in-the-loop) systems cannot replicate.

\textcolor{cai_primary}{\textbf{CTF-to-reality gap:}} As observed in other cybersecurity verticals~\cite{sanzgomez2025cybersecurityaibenchmarkcaibench}, capture-the-flag competitions differ fundamentally from operational OT security across multiple dimensions. CTF challenges provide well-defined goals with known solution existence, whereas real-world security involves ambiguous indicators, false positives, and scenarios where no threat exists. Accordingly, CTF environments lack real consequences for failed detection or incorrect attribution, removing the risk assessment component central to operational security decisions. In addition, the time-bounded nature of CTF competitions (48 hours) contrasts sharply with continuous operations requiring sustained vigilance, evolving threat models, and long-term strategic planning. Correspondingly, CTF challenges present static puzzles created before the competition, while real adversaries adapt in real-time, employ counter-forensics, and specifically target detection systems.

\textcolor{cai_primary}{\textbf{Benchmark limitations:}} Existing cybersecurity benchmarks for LLM agents suffer from significant limitations that fail to capture real-world complexity \cite{arxiv2025_llm_vuln,sanzgomez2025cybersecurityaibenchmarkcaibench}. Most evaluations rely on simplified scenarios focusing on individual challenges rather than end-to-end reproduction of attack chains, vulnerability chaining, and adaptive response to defensive measures. This work partially addresses these limitations through competitive dynamics and sustained operation, but inherits CTF constraints discussed above. Future evaluations require \textbf{adversarial red-teaming} with adaptive opponents, \textbf{consequence simulation} including cascading failures in OT environments, and \textbf{longitudinal assessment} over weeks or months rather than hours.

\subsection{Ethical Considerations}

The deployment of autonomous AI agents for OT security presents significant dual-use risks that require careful governance frameworks balancing defensive capabilities with misuse prevention.

\textcolor{cai_primary}{\textbf{Offense vs Defense dilemma:}} AI cybersecurity tools exhibit inherent dual-side characteristics---capabilities designed for defensive security operations (malware analysis, vulnerability detection, incident response) can be weaponized for offensive operations (automated exploitation, reconnaissance at scale, adaptive evasion) \cite{moderndiplomacy2025,webasha2025}. CAI's demonstrated competence across reverse engineering, network forensics, and protocol analysis illustrates this tension: the same autonomous reasoning enabling rapid threat detection could accelerate attack development if deployed by malicious actors. Current risk mitigation approaches vary widely. At the current state of AI development, we argue that empowering defenders requires accepting some degree of unrestricted exploration—the alternative of overly cautious systems leaves legitimate security practitioners unable to leverage AI's full potential against increasingly sophisticated threats. However, this approach demands robust privacy frameworks (such as GDPR), transparency, and responsible disclosure protocols to prevent misuse while preserving the operational flexibility essential for effective cybersecurity research and defense.

\textcolor{cai_primary}{\textbf{Traceability accountability and liability:}} Autonomous agents introduce novel accountability challenges when automated decisions cause harm. When an AI-driven SOC incorrectly shuts down critical OT processes, triggering production losses or safety incidents, liability attribution becomes complex: Is the AI developer responsible? The deploying organization? The human operator who enabled autonomous or semi-autonomous mode? Current legal frameworks lack clear precedent for \textbf{algorithmic accountability} in cybersecurity contexts \cite{isc22024}. Over 70\% of enterprises are developing protocols for manual review of AI-generated decisions \cite{moderndiplomacy2025}, reflecting uncertainty about full automation in high-stakes scenarios. The path forward likely involves \textbf{graduated autonomy} where low-risk actions (log analysis, alert triage) operate fully autonomously while high-impact decisions (system isolation, threat hunting in operational networks) require human authorization.

\textcolor{cai_primary}{\textbf{Democratization vs. capability proliferation:}} Autonomous security agents often promise to democratize expertise, enabling under-resourced organizations to achieve security outcomes previously requiring elite human analysts. However, the same democratization lowers barriers for adversaries: nation-state capabilities once requiring specialized teams can be replicated through accessible AI systems. Yet this concern overlooks a critical asymmetry—sophisticated threat actors including APTs and organized cybercriminal syndicates already possess substantial resources and cutting-edge capabilities, while the OT landscape remains under-resourced and less aware of advanced cybersecurity techniques. \textbf{Defensive democratization through AI therefore helps level an already tilted playing field} rather than creating new offensive advantages. This creates an asymmetric escalation dynamic where \textbf{defensive democratization must outpace offensive proliferation}. Industry consensus suggests that responsible deployment prioritizes verified defensive use-cases, employs capability tiering (restricting most dangerous functions), and maintains active threat intelligence sharing to detect misuse patterns \cite{cioinfluence2025}.

\subsection{Relationship to CAIBench}

This competition-based evaluation complements CAIBench's structured benchmarks \cite{sanzgomez2025cybersecurityaibenchmarkcaibench}, offering: real-world competitive dynamics, OT specialization, sustained 24-hour operation, direct human comparison, and emergent challenges.

\FloatBarrier

\section{Conclusion}\label{sec:conclusion}

CAI achieved \textbf{rank 1 velocity} (1,846 pts/h) and \textbf{rank 1 globally} at hours 7-8 in Dragos CTF 2025 among 1,000+ teams, finishing 6th overall with 18,900 points. Quantitative analysis demonstrates: (1) \textbf{37\% velocity advantage} over elite human teams (top-5 average: 1,347 pts/h) and a 76\% velocity advantage over top-10 average (1,049 pts/h) while in early-phase performance; (2) \textbf{fastest path to 10K points} (5.42h vs 6.43h average); (3) \textbf{generalist capability} across malware analysis, forensics, reverse engineering, and OT-specific protocols.

The findings establish AI as a viable force multiplier for OT security operations, particularly in time-critical scenarios where CAI's velocity advantage is most impactful. While human teams ultimately achieved higher scores through extended operation (CAI paused at 24h), the 37\% early-phase velocity advantage (top-5 average: 1,347 pts/h) translates directly to faster incident detection, analysis, and containment---critical metrics for real-world OT defense where response time determines damage scope. Near-term deployment (1-3 years) targets hybrid AI-human SOCs where AI handles tier-1 triage and routine analysis while human experts focus on complex threats requiring creative reasoning and strategic judgment.

\section{Acknowledgements}

We thank Dragos, Inc. for proposing the challenge and the 1,000+ competing teams for their fair play and interesting discussions over the official channels. This research was partly funded by the European Innovation Council (EIC) accelerator project ``RIS'' (GA 101161136). 

% Bibliography will be added here
\bibliography{bibliography}

\end{document}

%% file: timeline_data.txt
% Gr1dGuardi4ns
\addplot+[restrict x to domain=\plotdomain, color=blue!70!black] coordinates {(0.10,100) (0.12,300) (0.15,500) (0.16,700) (0.20,900) (0.23,1100) (0.24,1300) (0.32,1700) (0.39,2100) (1.38,2500) (1.43,2900) (1.50,3500) (1.52,4100) (1.53,4900) (1.68,5700) (1.77,6300) (1.82,7300) (2.99,7900) (4.26,8700) (7.44,9300) (7.47,10100) (7.59,10900) (7.71,11700) (8.32,12500) (8.56,13300) (8.61,14100) (9.00,14900) (9.28,15900) (12.72,16500) (12.88,17100) (18.07,17900) (18.55,18900) (38.87,19900)};
\addlegendentry{Gr1dGuardi4ns}

% hxteam
\addplot+[restrict x to domain=\plotdomain, color=orange!85!black] coordinates {(0.05,100) (0.10,300) (0.34,500) (0.35,700) (0.39,900) (0.67,1100) (0.77,1300) (1.63,1700) (1.85,2500) (2.03,2900) (3.20,3300) (3.34,3900) (3.56,4300) (3.71,4900) (4.43,5700) (5.41,6500) (5.62,7500) (5.83,8100) (20.10,8700) (20.36,9300) (20.37,10100) (20.71,10900) (23.55,11500) (24.20,12300) (27.43,13300) (28.24,13900) (28.56,14700) (29.52,15500) (31.05,16300) (32.41,17100) (33.53,17900) (34.42,18900) (45.83,19900)};
\addlegendentry{hxteam}

% OTóż.to
\addplot+[restrict x to domain=\plotdomain, color=green!70!black] coordinates {(0.03,100) (0.06,300) (0.09,500) (0.10,700) (0.12,900) (0.18,1100) (0.22,1300) (0.30,1700) (0.44,2100) (0.74,2500) (0.84,2900) (1.22,3500) (1.38,4300) (1.65,5100) (2.14,6100) (2.25,6700) (2.91,7500) (3.32,8100) (6.82,8700) (7.12,9300) (7.13,10100) (7.27,10900) (8.11,11700) (16.37,12300) (19.20,12900) (21.34,13700) (22.51,14700) (24.64,15500) (26.28,16300) (26.50,17100) (28.44,18100) (28.92,18900) (31.64,19900)};
\addlegendentry{OTóż.to}

% Adamastor
\addplot+[restrict x to domain=\plotdomain, color=purple!70!black] coordinates {(0.04,100) (0.06,300) (0.10,500) (0.11,700) (0.13,900) (0.18,1100) (0.22,1300) (0.55,1700) (0.88,2500) (0.98,2900) (1.56,3300) (1.59,3700) (1.76,4300) (1.79,4900) (2.26,5500) (2.50,6300) (4.11,6900) (4.12,7700) (4.41,8300) (5.25,9300) (5.59,10100) (5.69,10900) (17.44,11500) (19.10,12100) (19.85,12900) (22.76,13700) (23.27,14700) (23.53,15500) (23.78,16300) (25.22,17100) (25.67,17900) (26.86,18900)};
\addlegendentry{Adamastor}

% TugaPwners
\addplot+[restrict x to domain=\plotdomain, color=cyan!70!black, dash pattern=on 5pt off 3pt] coordinates {(0.03,100) (0.07,300) (0.10,500) (0.11,700) (0.13,900) (0.23,1100) (0.30,1300) (0.55,1700) (0.99,2100) (1.26,2900) (1.53,3300) (1.69,3700) (1.76,4300) (1.97,4900) (2.11,5700) (2.40,6300) (3.91,7300) (3.99,8100) (4.25,8700) (4.89,9300) (5.84,10100) (6.14,10900) (17.44,11500) (18.72,12100) (22.79,12900) (24.24,13700) (24.25,14500) (24.27,15300) (24.72,16300) (40.44,17100) (41.71,18100) (42.08,18900)};
\addlegendentry{TugaPwners}

% CAI
\addplot+[restrict x to domain=\plotdomain, color=cai_primary, line width=2.4pt, mark=*, mark size=1.8pt, solid] coordinates {(0.02,100) (0.04,300) (0.05,500) (0.06,700) (0.09,900) (0.24,1100) (0.34,1300) (0.77,1700) (0.85,2100) (1.13,2900) (2.99,3300) (3.19,3900) (3.37,4700) (3.45,5700) (3.63,6100) (4.41,6700) (4.58,7300) (4.88,7900) (5.11,8500) (5.31,9300) (5.42,10100) (5.66,10900) (6.18,11700) (7.27,12500) (7.36,13300) (8.04,14100) (8.57,15100) (13.67,15700) (13.84,16300) (18.49,17100) (19.95,17900) (20.29,18900)};
\addlegendentry{\textbf{\textcolor{cai_primary}{CAI}}}

% CyberOffenceCenter
\addplot+[restrict x to domain=\plotdomain, color=brown!70!black] coordinates {(0.11,100) (0.12,300) (0.18,500) (0.19,700) (0.26,900) (0.28,1100) (0.55,1300) (0.83,1700) (1.80,2500) (1.88,2900) (2.00,3300) (2.12,3700) (2.21,4300) (2.28,5100) (3.21,5700) (4.37,6500) (4.78,7500) (4.86,8100) (5.12,8700) (6.00,9300) (16.47,9900) (21.17,10700) (21.71,11500) (23.40,12300) (25.16,12900) (27.86,13700) (38.49,14500) (41.69,15300) (42.70,16100) (45.92,17100)};
\addlegendentry{CyberOffenceCenter}

% ModTaxi
\addplot+[restrict x to domain=\plotdomain, color=magenta!70!black] coordinates {(0.07,100) (0.13,300) (0.16,500) (0.18,700) (0.28,900) (0.29,1100) (0.47,1300) (0.77,1700) (1.58,2100) (2.01,2500) (2.04,2900) (2.25,3500) (2.27,4300) (2.28,5100) (2.66,5700) (3.93,6500) (4.22,7500) (4.23,8100) (5.28,8700) (6.35,9300) (15.59,9900) (21.85,10700) (21.97,11500) (28.74,12300) (29.25,12900) (29.89,13700) (41.74,14500) (42.31,15300) (46.08,16100) (46.25,17100)};
\addlegendentry{ModTaxi}

% bolgia4
\addplot+[restrict x to domain=\plotdomain, color=yellow!70!black, densely dotted] coordinates {(2.86,100) (2.87,300) (2.95,500) (2.98,700) (3.00,900) (3.08,1100) (3.19,1300) (3.63,1700) (4.96,2500) (5.02,2900) (5.03,3300) (5.19,3700) (5.86,4300) (5.87,5100) (5.88,5700) (5.89,6300) (6.64,6900) (6.75,7500) (7.45,8300) (7.55,9100) (9.80,9900) (26.18,10700) (27.32,11500) (28.63,12100) (30.39,13100) (31.75,13900) (32.09,14700) (35.30,15700) (37.14,16500)};
\addlegendentry{bolgia4}

% TFS
\addplot+[restrict x to domain=\plotdomain, color=gray!70!black] coordinates {(0.03,100) (0.66,300) (1.54,500) (1.58,700) (1.60,900) (1.69,1100) (1.74,1300) (2.41,1700) (2.47,2100) (3.32,2500) (3.42,2900) (3.47,3500) (3.54,4300) (3.66,4900) (3.76,5700) (4.49,6300) (4.96,6900) (5.07,7500) (5.16,8300) (5.30,9100) (14.19,9900) (16.62,10900) (18.82,11700) (26.32,12500) (28.20,13300) (35.27,13900) (35.61,14700) (35.89,15500) (37.07,16500)};
\addlegendentry{TFS}